\pdfoutput=1

\documentclass[english]{article}

\usepackage[T1]{fontenc}
\usepackage[utf8]{inputenc} % harmless on TL2023
\usepackage{lmodern}
\usepackage{xcolor}
\definecolor{HermesOrange}{HTML}{F37021}
\usepackage{microtype}
\usepackage{subcaption}
\usepackage[english]{babel}
\usepackage{algorithm}

\usepackage[left=0.55in,right=0.55in,top=0.4in,bottom=0.5in,includeheadfoot]{geometry}

\usepackage{graphicx}
\usepackage{float}
\usepackage{booktabs}
\usepackage{multirow}
\usepackage{siunitx}

\usepackage{amsmath,amssymb}
\usepackage{amsthm}
\newtheorem{theorem}{Theorem}
\theoremstyle{remark}
\newtheorem{remark}{Remark}
\theoremstyle{definition}

\theoremstyle{lemma}
\newtheorem{lemma}{Lemma}
\theoremstyle{corollary}
\newtheorem{corollary}{Corollary}
\theoremstyle{proposition}

\usepackage[numbers,sort&compress]{natbib}
\usepackage{animate}

\usepackage[hidelinks]{hyperref}

\makeatother
\title{The Coupled Scaled Dirichlet Distribution}
\author{
  Jared N. Lakhani\\
  \texttt{jaredlakhani@gmail.com}
}
\date{} % You can also write a fixed date or leave blank 
\begin{document}
% \newgeometry{left=0.65in,right=0.65in,top=0.4in,bottom=0.5in,includeheadfoot}
\maketitle
\begin{abstract}
We introduce a coupled scaled Dirichlet distribution for paired compositions that may have different dimensions. Building on the matched Gamma-sharing construction of \citet*{Trick2023} for ordinary Dirichlet vectors, we allow scaled Dirichlet margins and extend the construction to include crossed sharing between non-corresponding latent components. We propose a signed coupling parameter whose sign selects matched or crossed sharing and whose magnitude controls the strength of sharing. We then develop a time-varying extension in which observed covariates determine this parameter. The model can therefore switch between matched and crossed sharing over time while preserving both scaled Dirichlet margins. We derive the joint mixed moments and use them to estimate dependence after fitting the margins separately. The moments can be evaluated by Monte Carlo simulation or by two-dimensional numerical quadrature, regardless of the dimensions of the compositions. Simulations show that the estimator generally recovers the sharing pattern, although weak crossed sharing is harder to detect. In an application to four U.S. equity-index futures, the time-varying model modestly improved the held-out energy and variogram scores for the trading-volume and intraday price-range compositions relative to the static-dependence model.
\end{abstract}

\section{Introduction}
When two compositions are observed together, interest often lies in how the shares in one composition relate to those in the other. For example, across futures contracts, one may ask how the distribution of trading volume relates to the distribution of intraday price variation. Existing Gamma-sharing constructions provide a natural way to model such relationships. In particular, \citet{Trick2023} construct correlated ordinary Dirichlet vectors by allowing corresponding latent Gamma components to share common Gamma terms.\\

The present paper extends this idea in three directions. First, the ordinary Dirichlet margins are replaced by scaled-Dirichlet margins, allowing additional component-specific flexibility. Second, the matched sharing structure is generalised through a $K \times L$ matrix of shared Gamma components, permitting dependence between arbitrary components of the two compositions. A parsimonious signed one-parameter restriction is then proposed in which the sign of $\eta$ selects between matched and crossed sharing, while its magnitude controls the strength of the selected sharing pattern. Third, the static dependence parameterisation is extended to allow the dependence structure to vary over time through a time-varying parameter $\eta_t$, thereby allowing both the strength and pattern of dependence to evolve across observations.\\

To estimate dependence under either specification, we derive the mixed moments of the joint distribution and show how to evaluate them by Monte Carlo simulation or two-dimensional quadrature. After fitting the two margins separately, we use the resulting cross-moments to estimate the dependence parameters from the paired observations. We evaluate this procedure in simulations and apply it to daily trading-volume and price-range shares for four U.S. equity-index futures contracts, where we examine whether the relationship between the two compositions varies with market conditions.

\section{The Coupled Scaled Dirichlet Distribution}
The construction developed below builds directly on the Gamma-sharing approach
of \citet{Trick2023}, but differs by allowing scaled-Dirichlet margins and a
general cross-component sharing matrix.\\

Let $\mathbf{X}_i
=
(X_{i1},\ldots,X_{iK})^{\top}
\in
\mathcal{S}^{K-1}$
and $\mathbf{Y}_i
=
(Y_{i1},\ldots,Y_{iL})^{\top}
\in
\mathcal{S}^{L-1}$ for $i=1,\ldots,n,$
denote a pair of compositional random vectors, where the simplex $\mathcal{S}^{d-1}
=
\left\{
\mathbf{x}\in(0,1)^d:
\sum_{j=1}^{d}x_j=1
\right\}$. The proposed coupled model is constructed from three collections of mutually
independent Gamma random variables. Let $\mathbf{a}
=
(a_1,\ldots,a_K)^{\top}$ and $\mathbf{b}
=
(b_1,\ldots,b_L)^{\top}$
denote the marginal shape parameters, and let $\boldsymbol{\Delta}
=
(\delta_{k\ell})_
{\substack{k=1,\ldots,K; \ \ell=1,\ldots,L}}$
be a non-negative $K\times L$ matrix governing dependence between the
two compositions.\\

Now, define the row and column sums of $\boldsymbol{\Delta}$ as $\delta_{k+}
=
\sum_{\ell=1}^{L}\delta_{k\ell}$ for $
k=1,\ldots,K$
and $\delta_{+\ell}
=
\sum_{k=1}^{K}\delta_{k\ell}$ for $\ell=1,\ldots,L$. \\
For each observation $i$, we define the private Gamma variables as
\[
A_{ik}
\sim
\operatorname{Gamma}
\left(
a_k-\delta_{k+},1
\right),
\qquad
k=1,\ldots,K,
\]
and
\[
B_{i\ell}
\sim
\operatorname{Gamma}
\left(
b_\ell-\delta_{+\ell},1
\right),
\qquad
\ell=1,\ldots,L,
\]
together with the shared Gamma variables
\[
D_{ik\ell}
\sim
\operatorname{Gamma}
\left(
\delta_{k\ell},1
\right),
\qquad
k=1,\ldots,K,
\quad
\ell=1,\ldots,L.
\]

All $A_{ik}$, $B_{i\ell}$ and $D_{ik\ell}$ are assumed mutually
independent. When $\delta_{k\ell}=0$, we set $D_{ik\ell}=0$ with probability one
and omit this coordinate and the corresponding row of $\mathbf H$
when evaluating the Dirichlet average in Theorem~1. Naturally, the parameter space is restricted such that $a_k>\delta_{k+}$ for
$k=1,\ldots,K$ and $b_\ell>\delta_{+\ell}$ for $\ell=1,\ldots,L$.\\

The variables underlying the two compositions (the latent Gamma variables) are then
defined as
\[
U_{ik}
=
A_{ik}
+
\sum_{\ell=1}^{L}D_{ik\ell},
\qquad
k=1,\ldots,K,
\]
and
\[
V_{i\ell}
=
B_{i\ell}
+
\sum_{k=1}^{K}D_{ik\ell},
\qquad
\ell=1,\ldots,L.
\]

Since independent Gamma random variables with a common rate parameter
are additive, we have
\[
U_{ik}
\overset{\mathrm{ind}}{\sim}
\operatorname{Gamma}(a_k,1),
\qquad
k=1,\ldots,K,
\]
and
\[
V_{i\ell}
\overset{\mathrm{ind}}{\sim}
\operatorname{Gamma}(b_\ell,1),
\qquad
\ell=1,\ldots,L.
\]
The components of $\mathbf U_i$ are mutually independent, as are the
components of $\mathbf V_i$. However, $\mathbf U_i$ and $\mathbf V_i$
are jointly dependent, since $U_{ik}$ and $V_{i\ell}$ share the common
Gamma variable $D_{ik\ell}$. In particular, $\operatorname{Cov}(U_{ik},V_{i\ell})
=
\delta_{k\ell}$.\\

Now, let $\boldsymbol{\rho}
=
(\rho_1,\ldots,\rho_K)^{\top}$
and $\boldsymbol{\kappa}
=
(\kappa_1,\ldots,\kappa_L)^{\top}$
be strictly positive scale vectors with references $\rho_K = 1$ and $\kappa_L = 1$. The observed compositions are then
obtained by
\[
X_{ik}
=
\frac{
\rho_k U_{ik}
}{
\displaystyle
\sum_{h=1}^{K}\rho_hU_{ih}
},
\qquad
k=1,\ldots,K,
\]
and
\[
Y_{i\ell}
=
\frac{
\kappa_\ell V_{i\ell}
}{
\displaystyle
\sum_{m=1}^{L}\kappa_mV_{im}
},
\qquad
\ell=1,\ldots,L.
\]
This implies the marginal distributions
\[
\mathbf X_i
\sim
\operatorname{SDirichlet}_{K}
(\mathbf a,\boldsymbol\rho)
\]
and
\[
\mathbf Y_i
\sim
\operatorname{SDirichlet}_{L}
(\mathbf b,\boldsymbol\kappa).
\]
Thus, the proposed construction preserves scaled Dirichlet marginal
distributions for both compositions, while dependence between
$\mathbf X_i$ and $\mathbf Y_i$ is induced through the shared Gamma
variables governed by $\boldsymbol{\Delta}$.

\begin{theorem}[Mixed moments of the coupled scaled
Dirichlet distribution]
Let $(\mathbf X,\mathbf Y)$
follow the coupled scaled Dirichlet construction, and let $\mathbf p=(p_1,\ldots,p_K)^\top$ and $\mathbf q=(q_1,\ldots,q_L)^\top$, where $p_k,q_\ell\in\mathbb N_0$. Define $P=\sum_{k=1}^{K}p_k$ and $
Q=\sum_{\ell=1}^{L}q_\ell$, and let $\boldsymbol\gamma
=
\left(
\mathbf a-\boldsymbol\Delta\mathbf 1_L,\;
\mathbf b-\boldsymbol\Delta^\top\mathbf 1_K,\;
\operatorname{vec}(\boldsymbol\Delta)
\right)^\top \in \mathbb{R}^M$, for which $\mathbf Z\sim
\operatorname{Dirichlet}_M(\boldsymbol\gamma)$. \\

Now, given a stacked mutually independent Gamma vector $\mathbf G
=
(
A_1,\ldots,A_K,
B_1,\ldots,B_L,
D_{11},\ldots,D_{KL}
)^\top$: for each $k=1,\ldots,K$, let
$\mathbf c_k^X\in\{0,1\}^{M}$ denote the incidence vector having
entries equal to one in the positions corresponding to
$A_k,D_{k1},\ldots,D_{kL}$, and zero elsewhere. Similarly, for each
$\ell=1,\ldots,L$, let
$\mathbf c_\ell^Y\in\{0,1\}^{M}$ denote the incidence vector having
entries equal to one in the positions corresponding to
$B_\ell,D_{1\ell},\ldots,D_{K\ell}$, and zero elsewhere.\\

Now define
\[
\mathbf H
=
\left[
\mathbf c_1^X,\ldots,\mathbf c_K^X,\;
\mathbf c_1^Y,\ldots,\mathbf c_L^Y,\;
\mathbf d_X,\mathbf d_Y
\right],
\]
where $\mathbf d_X
=
\sum_{k=1}^{K}\rho_k\mathbf c_k^X$ and $
\mathbf d_Y
=
\sum_{\ell=1}^{L}\kappa_\ell\mathbf c_\ell^Y$.
Further, let $\boldsymbol\nu
=
\left(
p_1,\ldots,p_K,\;
q_1,\ldots,q_L,\;
-P,-Q
\right)^\top$. Then
\begin{align}
\boxed{
\begin{aligned}
\mathbb E
\left[
\prod_{k=1}^{K}X_k^{p_k}
\prod_{\ell=1}^{L}Y_\ell^{q_\ell}
\right]
=
\left(
\prod_{k=1}^{K}\rho_k^{p_k}
\right)
\left(
\prod_{\ell=1}^{L}\kappa_\ell^{q_\ell}
\right) \times
\mathcal R
\left(
\boldsymbol\gamma;
\mathbf H;
\boldsymbol\nu
\right), \label{eq: e_mixed}
\end{aligned}
}
\end{align}
where $\mathcal R
\left(
\boldsymbol\gamma;
\mathbf H;
\boldsymbol\nu
\right)
=
\mathbb E_{\mathbf Z\sim
\operatorname{Dirichlet}(\boldsymbol\gamma)}
\left[
\prod_{r}
(\mathbf h_r^\top\mathbf Z)^{\nu_r}
\right]$
is the corresponding Carlson multiple $R$-Dirichlet average given in Lemma \ref{lem:carlson_multiple_R}.
\end{theorem}

\begin{proof}
Stack the mutually independent Gamma variables as
\[
\mathbf G
=
(
A_1,\ldots,A_K,
B_1,\ldots,B_L,
D_{11},\ldots,D_{KL}
)^\top
\]
and define
\[
T=\sum_rG_r,
\qquad
\mathbf Z=\frac{\mathbf G}{T}.
\]
Since the elements of $\mathbf G$ are independent Gamma random
variables with common rate parameter $1$,
\[
\mathbf Z
\sim
\operatorname{Dirichlet}_{M}(\boldsymbol\gamma), 
\]
where $\boldsymbol\gamma
=
\left(
\mathbf a-\boldsymbol\Delta\mathbf 1_L,\;
\mathbf b-\boldsymbol\Delta^\top\mathbf 1_K,\;
\operatorname{vec}(\boldsymbol\Delta)
\right)^\top \in \mathbb{R}^{M}$ with $M = K+L+KL$.\\

% Now since $G_r = TZ_r$, that is, each Gamma variable equals $T$ times its corresponding $Z$- component, we have $A_k = TZ_{A_k}$ and $D_{kl} = T Z_{D_{kl}}$. Hence
Now we have
\begin{align}
    U_k &= A_k + \sum_{\ell=1}^{L}D_{k\ell} \nonumber\\
    & = (\mathbf c_k^X)^\top\mathbf G \nonumber \\ 
    & = T(\mathbf c_k^X)^\top\mathbf Z \nonumber
\end{align}
% \begin{align}
%     U_k &= A_k + \sum_{\ell=1}^{L}D_{k\ell} \nonumber \\
%     & =  T\left( Z_{A_k} +  \sum_{\ell=1}^{L}Z_{k\ell} \right) \nonumber \\
%     & = T(\mathbf c_k^X)^\top\mathbf Z .\nonumber
% \end{align}
Similarly, $V_\ell
=
T(\mathbf c_\ell^Y)^\top\mathbf Z$. Additionally, we have
\begin{align}
    \sum_{h = 1}^{K} \rho_h U_h& =  \sum_{h = 1}^{K} \rho_h  T(\mathbf c_h^X)^\top\mathbf Z \nonumber \\
    & = T \mathbf d_X^\top \mathbf{Z}, \nonumber
\end{align}
and similarly, $\sum_{m = 1}^{L} \kappa_m V_m = T \mathbf d_Y^\top \mathbf{Z}$.
Consequently,
\[
X_k
=
\rho_k
\frac{
(\mathbf c_k^X)^\top\mathbf Z
}{
\mathbf d_X^\top\mathbf Z
}
\]
and
\[
Y_\ell
=
\kappa_\ell
\frac{
(\mathbf c_\ell^Y)^\top\mathbf Z
}{
\mathbf d_Y^\top\mathbf Z
}.
\]
It follows that
\begin{align*}
\prod_{k=1}^{K}X_k^{p_k}
\prod_{\ell=1}^{L}Y_\ell^{q_\ell}
={}&
\left(
\prod_{k=1}^{K}\rho_k^{p_k}
\right)
\left(
\prod_{\ell=1}^{L}\kappa_\ell^{q_\ell}
\right)
\\
&\times
\prod_{k=1}^{K}
[(\mathbf c_k^X)^\top\mathbf Z]^{p_k}
\prod_{\ell=1}^{L}
[(\mathbf c_\ell^Y)^\top\mathbf Z]^{q_\ell}
\\
&\times
(\mathbf d_X^\top\mathbf Z)^{-P}
(\mathbf d_Y^\top\mathbf Z)^{-Q}.
\end{align*}
Taking expectations with respect to
$\mathbf Z\sim\operatorname{Dirichlet}(\boldsymbol\gamma)$
and applying the Dirichlet-average representation of Carlson's
multiple $R$-function in Lemma \ref{lem:carlson_multiple_R} gives the result.
\end{proof}

\begin{corollary}[Cross-moments]
For $j=1,\ldots,K$ and $\ell=1,\ldots,L$,
\begin{align}
    \boxed{
\mathbb E[X_jY_\ell]
=
\rho_j\kappa_\ell
\mathcal R
\left(
\boldsymbol\gamma;
\left[
\mathbf c_j^X,
\mathbf c_\ell^Y,
\mathbf d_X,
\mathbf d_Y
\right];
(1,1,-1,-1)^\top
\right). }\label{eq: e_xy}
\end{align}
\end{corollary}

\subsection{A parsimonious signed parameterisation of $\boldsymbol{\Delta}$}
\label{subsec:signed_eta}

The matrix $\boldsymbol{\Delta}=(\delta_{k\ell})$ describes dependence between
the latent Gamma vectors, with
$\delta_{k\ell}=\operatorname{Cov}(U_k,V_\ell)$. An unrestricted
$K\times L$ matrix introduces up to $KL$ dependence parameters, subject to the
Gamma shape constraints. Such a parameterisation may be difficult to identify
empirically because different latent sharing configurations can generate similar
observable mixed moments.\\

Our parameterisation builds on, but differs from, existing Gamma-sharing
constructions. \citet{Trick2023} generate dependence between Dirichlet vectors
by allowing corresponding latent Gamma components to share a common Gamma
term. In this construction, sharing is \emph{matched}: the shared component
associated with coordinate $k$ enters both $U_k$ and $V_k$. A parsimonious
version of this idea restricts the amount of sharing relative to its
component-specific upper bound, $\min(a_k,b_k)$, thereby allowing the strength
of matched sharing to be represented by a single scalar parameter.
Separately, the constructions of \citet{ArnoldNg2011,ArnoldArvanitis2021}
show that more general Gamma-sharing schemes, including cross-component
sharing, can generate a broader range of dependence patterns after
normalisation.\\

The contribution of the present paper is to combine these ideas in a single
parsimonious signed parameterisation that is applicable to paired
compositions of arbitrary dimensions $K$ and $L$. For $\eta>0$, we retain the
proportional matched-sharing structure, so that corresponding latent
components share Gamma mass. For $\eta<0$, we introduce a new
\emph{crossed-sharing} rule in which Gamma mass is shared between
non-corresponding components. The sign of $\eta$ therefore selects the
sharing topology, matched or crossed, while $|\eta|$ controls its strength.
Importantly, a negative value of $\eta$ does not imply a negative latent
Gamma covariance: all shared-Gamma covariances remain non-negative. Rather,
crossed sharing can induce negative associations between observed
compositional components after normalisation.\\

The proposed restriction therefore extends the matched-sharing construction
to a signed, one-parameter dependence model that accommodates both matched
and crossed latent sharing while preserving the scaled-Dirichlet margins.
The normalisation by $\max(K,L)-1$ on the crossed branch is introduced to
ensure admissibility of the latent Gamma shape parameters for general,
including rectangular, $K\times L$ settings. Specifically, for $K,L\geq2$ and $a_k,b_\ell>0$, define
\begin{equation}
\delta_{k\ell}(\eta)=
\begin{cases}
\eta\min(a_k,b_\ell),
& \eta\geq0,\quad k=\ell,\\[4pt]
\displaystyle\frac{|\eta|}{\max(K,L)-1}\min(a_k,b_\ell),
& \eta<0,\quad k\neq\ell,\\[7pt]
0,&\text{otherwise},
\end{cases}
\qquad -1<\eta<1.
\end{equation}

The same $\eta$ applies to all paired observations. Positive values specify \emph{matched sharing}, linking corresponding latent components; negative values specify \emph{crossed sharing}, linking different components. Its magnitude controls the amount of sharing within the selected pattern. Thus, the sign selects the pattern and the magnitude selects its strength, without estimating separate parameters for the two patterns. When $K\neq L$, components beyond the first $\min(K,L)$ have no matched partner and receive no shared contribution on the positive branch.\\

For $\eta\geq0$, each matched pair uses the fraction $\eta$ of its maximum feasible shared shape, $\min(a_k,b_k)$. For $\eta<0$, sharing is distributed across multiple pairs. The divisor $\max(K,L)-1$ accounts for the largest number of crossed links in any row or column. When $K=L$, it reduces to $K-1$. In the two-part case, crossed sharing is equivalent to matched sharing after reversing one composition's component labels; with three or more components on each side, crossed sharing links each component to several others. \\

Now, each row and column contains at most one matched link or at most $\max(K,L)-1$ crossed links. Since $\min(a_k,b_\ell)\leq a_k$ and $\min(a_k,b_\ell)\leq b_\ell$, both branches satisfy
\begin{equation}
0\leq\delta_{k+}\leq|\eta|a_k<a_k,
\qquad
0\leq\delta_{+\ell}\leq|\eta|b_\ell<b_\ell. \nonumber
% \label{eq:signed_eta_budgets}
\end{equation}
Hence all private Gamma shapes remain strictly positive throughout the fixed interval $\eta \in (-1,1)$, regardless of the marginal shapes. Specifically, take mutually independent Gamma variables $A_k\sim\operatorname{Gamma}(a_k-\delta_{k+},1)$, $B_\ell\sim\operatorname{Gamma}(b_\ell-\delta_{+\ell},1)$ and $
D_{k\ell}\sim\operatorname{Gamma}(\delta_{k\ell},1)$, and define $U_k=A_k+\sum_{\ell=1}^{L}D_{k\ell}$ and $V_\ell=B_\ell+\sum_{k=1}^{K}D_{k\ell}$. Since independent Gamma variables with a common rate parameter are additive, we have $U_k\sim\operatorname{Gamma}(a_k,1)$ and $V_\ell\sim\operatorname{Gamma}(b_\ell,1)$. The normalised compositions therefore retain the exact margins $\mathbf{X}\sim\operatorname{SDirichlet}_K(\mathbf{a},\boldsymbol{\rho})$ and $
\mathbf{Y}\sim\operatorname{SDirichlet}_L(\mathbf{b},\boldsymbol{\kappa})$. At $\eta=0$, all shared variables vanish, so $\mathbf{U}$ and $\mathbf{V}$, and hence $\mathbf{X}$ and $\mathbf{Y}$, are independent.\\

Furthermore, the parameter $\eta$ is a sharing parameter, not a correlation coefficient. In particular,
\[
\operatorname{Corr}(U_k,V_\ell)
=\frac{\delta_{k\ell}(\eta)}{\sqrt{a_kb_\ell}}
\geq0
\]
on both branches: negative $\eta$ changes which components share Gamma variation, rather than introducing negative latent covariances. Observed covariances depend additionally on $\mathbf{a},\mathbf{b},\boldsymbol{\rho},\boldsymbol{\kappa}$ through normalisation. Thus, $\delta_{k\ell}=0$ need not imply $\operatorname{Cov}(X_k,Y_\ell)=0$, and negative $\eta$ does not generally imply negative covariance between every pair of corresponding observed components.

\subsection{A time-varying extension of the signed parameterisation of $\mathbf{\Delta}$}
\label{sec:dynamic_eta}

The signed parameterisation introduced in the previous section assumes that
the dependence parameter $\eta$ is constant across all observations. This is
appropriate when the dependence structure between the two compositions is
assumed to remain constant over the sample period. In many applications,
however, the strength or form of dependence may change with observable
conditions. We therefore extend the model by allowing the signed dependence
parameter, $\eta$, to vary across observations.\\

For observation $t$, let
\[
\eta_t
=
\tanh\left(
\beta_0+\mathbf{q}_t^\top\boldsymbol{\beta}
\right),
\]
where $\mathbf{q}_t$ is a $P\times 1$ vector of observed state variables,
$\beta_0$ is an intercept, and $\boldsymbol{\beta}$ is a corresponding
$P\times 1$ vector of coefficients. The hyperbolic tangent transformation
ensures that $-1<\eta_t<1$ for every observation, so that the same parameter restrictions established
for the static signed model continue to hold.\\

The dependence matrix is now observation-specific and is denoted by $\Delta_t=\Delta(\eta_t)$. Its elements are given by
\[
\delta_{k\ell,t}
=
\begin{cases}
\eta_t \min(a_k,b_\ell),
&
\eta_t\geq 0,\quad k=\ell,
\\[6pt]
\dfrac{|\eta_t|}{\max(K,L)-1}
\min(a_k,b_\ell),
&
\eta_t<0,\quad k\neq\ell,
\\[8pt]
0,
&
\text{otherwise}.
\end{cases}
\]

The interpretation of $\eta_t$ is therefore unchanged from the static model.
Its sign determines the form of latent Gamma sharing, while its magnitude
determines the strength of that sharing. When $\eta_t>0$, corresponding
components are linked through matched shared-Gamma variables. When
$\eta_t<0$, sharing occurs between non-matched components. When
$\eta_t=0$, all shared-Gamma terms vanish and the two latent Gamma vectors
are independent.\\

The important difference is that the sharing structure is now allowed to
change with $\mathbf{q}_t$. For example, a change in the state variables may move $\eta_t$ from a weakly positive value to a strongly
positive value, thereby increasing the amount of matched sharing. Similarly, changes in the state variables may cause \(\eta_t\) to cross zero. A change from a positive to a negative value moves the model from matched to crossed sharing, while a change from a negative to a positive value moves it from crossed to matched sharing.\\

Now as before, for each observation, define $A_{tk}
\sim
\operatorname{Gamma}
\left(
a_k-\delta_{k+,t},1
\right)$ for $k=1,\ldots,K$
and $B_{t\ell}
\sim
\operatorname{Gamma}
\left(
b_\ell-\delta_{+\ell,t},1
\right)$ for $
\ell=1,\ldots,L$
together with $D_{tk\ell}
\sim
\operatorname{Gamma}
\left(
\delta_{k\ell,t},1
\right)$
where $\delta_{k+,t}
=
\sum_{\ell=1}^{L}\delta_{k\ell,t}$ and $
\delta_{+\ell,t}
=
\sum_{k=1}^{K}\delta_{k\ell,t}$. The latent Gamma variables are then $U_{tk}
=
A_{tk}
+
\sum_{\ell=1}^{L}D_{tk\ell}$ and $V_{t\ell}
=
B_{t\ell}
+
\sum_{k=1}^{K}D_{tk\ell}$. Since $|\eta_t|<1$, the same row- and column-sum arguments used for the
static model ensure that all private Gamma shape parameters remain positive.
Furthermore, $U_{tk}\sim\operatorname{Gamma}(a_k,1)$ and $
V_{t\ell}\sim\operatorname{Gamma}(b_\ell,1)$
for every $t$. Consequently, conditional on the state variables,
\[
\mathbf{X}_t\mid\mathbf{q}_t
\sim
\operatorname{SDirichlet}_K(\mathbf{a},\boldsymbol{\rho}),
\]
and
\[
\mathbf{Y}_t\mid\mathbf{q}_t
\sim
\operatorname{SDirichlet}_L(\mathbf{b},\boldsymbol{\kappa}).
\]
Thus, the extension allows the cross-compositional dependence structure to
change over time without changing the specified marginal scaled-Dirichlet
distributions. The parameters
$\mathbf{a}$, $\boldsymbol{\rho}$, $\mathbf{b}$ and
$\boldsymbol{\kappa}$ continue to describe the marginal behaviour of the two
compositions, while $\eta_t$ determines the observation-specific latent
sharing structure.\\

It is useful to emphasise that $\eta_t$ remains a signed sharing parameter
rather than a correlation coefficient. In particular, a negative value of
$\eta_t$ does not imply that all observed cross-correlations are negative.
Instead, the sign of $\eta_t$ determines which latent Gamma components share
variation, while the resulting dependence between the observed
compositional components also depends on the marginal parameters and on the
normalisation used to construct $\mathbf{X}_t$ and $\mathbf{Y}_t$.

\section{Estimation}
\label{sec:estimation}

The parameters of the coupled scaled Dirichlet distribution are
estimated using a two-stage method-of-moments approach. The first
stage estimates the parameters governing the two marginal scaled
Dirichlet distributions, while the second stage estimates the
dependence parameters contained in $\boldsymbol{\Delta}$ using mixed
moments of the paired compositions.\\

Let $\left\{
(\mathbf X_i,\mathbf Y_i)
\right\}_{i=1}^{N}$
denote an independent sample from the coupled scaled Dirichlet
distribution. The parameter vector is $\boldsymbol{\theta}
=
\left(
\mathbf a^\top,
\boldsymbol\rho^\top,
\mathbf b^\top,
\boldsymbol\kappa^\top,
\operatorname{vec}(\boldsymbol\Delta)^\top
\right)^\top$, 
subject to the reference constraints $\rho_K=1$ and $
\kappa_L=1$.

\subsection{Estimation of the marginal parameters} \label{sec:marginal_estimation}

By construction, the marginal distributions are $\mathbf X_i
\sim
\operatorname{SDirichlet}_{K}
(\mathbf a,\boldsymbol\rho)$
and $\mathbf Y_i
\sim
\operatorname{SDirichlet}_{L}
(\mathbf b,\boldsymbol\kappa)$. Importantly, these marginal distributions do not depend on
$\boldsymbol\Delta$. Hence, the marginal parameters: ${\mathbf a}$, $
{\boldsymbol\rho}$, ${\mathbf b}$, and ${\boldsymbol\kappa}$, may first be
estimated separately using the scaled Dirichlet moments derived in \citet{lakhani2026scaleddirichletdistributionlauricella}.

\subsection{Estimation of the static dependence parameter}

Conditional on the first-stage estimates
$\widehat{\mathbf a}$,
$\widehat{\boldsymbol\rho}$,
$\widehat{\mathbf b}$, and
$\widehat{\boldsymbol\kappa}$,
the only remaining unknown parameter is the static dependence parameter
$\eta$. Once $\eta$ is specified, the entire matrix $\boldsymbol{\Delta}(\eta)$ is therefore determined.\\

The marginal distributions of $\mathbf X_i$ and $\mathbf Y_i$ do not depend on $\eta$. Consequently, marginal moments such as $\mathbb{E}(X_k)$ and $\mathbb{E}(Y_\ell)$ contain no information about the dependence parameter. Estimation of $\eta$ therefore relies on mixed moments involving components from both compositions, such as $\mathbb{E}(X_kY_\ell)$ from Equation~\eqref{eq: e_xy}.\\

Let
$\mathbf p=(p_1,\ldots,p_K)^\top\in\mathbb N_0^K$
and
$\mathbf q=(q_1,\ldots,q_L)^\top\in\mathbb N_0^L$.
For a selected pair $(\mathbf p,\mathbf q)$, define the theoretical mixed moment from Equation~\eqref{eq: e_mixed} as

\begin{equation}
m_{\mathbf p,\mathbf q}(\eta)
=
\left(
\prod_{k=1}^{K}\rho_k^{p_k}
\right)
\left(
\prod_{\ell=1}^{L}\kappa_\ell^{q_\ell}
\right)
\mathcal R
\left(
\boldsymbol\gamma\left(\boldsymbol{\Delta}(\eta)\right);
\mathbf H;
\boldsymbol\nu_{\mathbf p,\mathbf q}
\right), \nonumber
\end{equation}
where $\boldsymbol\gamma\left(\boldsymbol{\Delta}(\eta)\right)=
\begin{pmatrix}
\mathbf a-\boldsymbol{\Delta}(\eta)\mathbf 1_L, 
\mathbf b-\boldsymbol{\Delta}(\eta)^\top\mathbf 1_K, \operatorname{vec}\left(\boldsymbol{\Delta}(\eta)\right)
\end{pmatrix}^\top$. The corresponding sample mixed moment is
\begin{equation}
\widehat m_{\mathbf p,\mathbf q}=
\frac{1}{N}
\sum_{i=1}^{N}
\left(
\prod_{k=1}^{K}X_{ik}^{p_k}
\right)
\left(
\prod_{\ell=1}^{L}Y_{i\ell}^{q_\ell}
\right). \nonumber
\end{equation}

In the second stage, the marginal parameters entering the theoretical mixed moments are replaced by their first-stage estimates. Hence, $m_{\mathbf p,\mathbf q}
\left(
\eta;
\widehat{\mathbf a},
\widehat{\boldsymbol\rho},
\widehat{\mathbf b},
\widehat{\boldsymbol\kappa}
\right)$ is a function of $\eta$ alone.\\

Because the dependence structure contains only one unknown scalar parameter, it is not necessary to estimate the individual elements of $\boldsymbol{\Delta}$ separately. In principle, one sufficiently informative mixed-moment condition is sufficient to identify $\eta$. Second-order cross-moments of the form $\mathbb{E}(X_kY_\ell)$ provide a natural set of dependence moment conditions. Although there are $KL$ possible second-order cross-moments, the simplex restrictions
$\sum_{k=1}^{K}X_k=1$
and
$\sum_{\ell=1}^{L}Y_\ell=1$
imply that only $(K-1)(L-1)$ contain independent cross-dependence information.\\

Let $\widehat{\mathbf m}=
\left(
\widehat m_{\mathbf p^{(s)},\mathbf q^{(s)}}
\right)_{s=1}^{S}$ denote the vector of selected sample mixed moments, and let $\mathbf m(\eta)=
\left(
m_{\mathbf p^{(s)},\mathbf q^{(s)}}(\eta)
\right)_{s=1}^{S}$ denote the corresponding vector of theoretical mixed moments evaluated using the first-stage marginal parameter estimates. The dependence parameter is estimated by minimising

\begin{equation}
\widehat{\eta}=
\arg\min_{\eta\in\Omega}
\left[
\widehat{\mathbf m} -
\mathbf m(\eta)
\right]^\top
\mathbf W
\left[
\widehat{\mathbf m}- 
\mathbf m(\eta)
\right],
\label{eq: eta_gmm}
\end{equation}
where $\mathbf W$ is a positive definite weighting matrix. If a single mixed-moment condition is used, so that $S=1$, the second-stage estimator is exactly identified. If $S>1$, the model is overidentified and the additional mixed moments allow a broader set of cross-compositional dependence information to contribute to the estimation of $\eta$. In the simplest implementation, $\mathbf W$ may be taken to be the identity matrix, although alternative weighting matrices may be used.\\

The static parameterisation is continuous at $\eta=0$ but has different dependence structures on either side of zero. Estimation may therefore be carried out separately over the crossed branch, $\eta\in(-1,0)$, and the matched branch, $\eta\in(0,1)$, together with the independence case $\eta=0$. The estimated value is selected as the candidate producing the smallest value of the objective function in Equation~\eqref{eq: eta_gmm}. This branchwise procedure avoids numerical optimisation across the change in dependence structure at zero.\\

The admissible parameter space follows from the shared-Gamma construction. For the matched branch, each component participates in at most one shared-Gamma term, so that $\sum_{\ell=1}^{L}\delta_{k\ell}(\eta)
\leq
|\eta|a_k
<a_k$ and $\sum_{k=1}^{K}\delta_{k\ell}(\eta)
\leq
|\eta|b_\ell
<b_\ell$. For the crossed branch, each row or column contains at most
$\max(K,L)-1$ non-zero shared-Gamma terms. Hence, $\sum_{\ell=1}^{L}\delta_{k\ell}(\eta)
\leq
\frac{|\eta|}{\max(K,L)-1}
\left[\max(K,L)-1\right]a_k=
|\eta|a_k
<a_k$, and similarly, $\sum_{k=1}^{K}\delta_{k\ell}(\eta)
<
b_\ell$. Therefore, the private Gamma shape parameters remain positive for all components whenever $|\eta|<1$. The admissible parameter space is consequently $\Omega = 
\eta:-1<\eta<1$.\\

Finally, the estimated dependence matrix is obtained directly from $\widehat{\eta}$ as:
\begin{equation}
\widehat{\delta}_{k\ell}
=
\begin{cases}
\widehat{\eta}\min(\widehat a_k,\widehat b_\ell),
& \widehat{\eta}\geq0, k=\ell,\\
\dfrac{|\widehat{\eta}|}{\max(K,L)-1}
\min(\widehat a_k,\widehat b_\ell),
& \widehat{\eta}<0, k\neq\ell,\\
0,
& \text{otherwise}.
\end{cases} \nonumber
\end{equation}
Thus, the sign of $\widehat{\eta}$ determines whether the fitted dependence structure is matched or crossed, while $|\widehat{\eta}|$ determines the strength of the shared-Gamma coupling.

\subsection{Estimation of the time-varying dependence parameters}
\label{sec:dynamic_estimation}

The time-varying extension introduces the parameter vector $\boldsymbol{\theta}
=
\left(
\beta_0,
\boldsymbol{\beta}^\top
\right)^\top$, which determines the observation-specific dependence parameter $\eta_t(\boldsymbol{\theta})
=
\tanh\left(
\beta_0+\mathbf{q}_t^\top\boldsymbol{\beta}
\right)$. As in the static model, estimation proceeds in two stages. In the first
stage, the marginal scaled-Dirichlet parameters $\mathbf{a},
\boldsymbol{\rho},
\mathbf{b}$ and $
\boldsymbol{\kappa}$ are estimated separately using the marginal moment conditions described in
Section~\ref{sec:marginal_estimation}. These estimates are then treated as
fixed when estimating the dynamic dependence parameters.\\

Now, let $\mathbf{F}_t
=
\left(
F_{1t},
\ldots,
F_{St}
\right)^\top$ denote the vector of observed cross-moment quantities used for dependence
estimation. For second-order cross-moments, an individual element takes the
form $F_{st}=X_{tk_s}Y_{t\ell_s}$. The corresponding model-implied moment is $m_s\left(\eta_t\right)
=
E_{\eta_t}
\left[
X_{k_s}Y_{\ell_s}
\right]$, where the expectation is evaluated using the first-stage marginal parameter
estimates. Collecting the $S$ theoretical moments gives $\mathbf{m}(\eta_t)
=
\left(
m_1(\eta_t),
\ldots,
m_S(\eta_t)
\right)^\top$. The observation-specific moment residual, that is, the difference between the observed cross-products and their model-implied
values, is therefore
\[
\mathbf{e}_t(\boldsymbol{\theta})
=
\mathbf{F}_t
-
\mathbf{m}
\left(
\eta_t(\boldsymbol{\theta})
\right).
\]
If the dynamic dependence specification is appropriate, these differences
should not display systematic patterns. In particular, they should be close
to zero on average and should not systematically increase or decrease with
the state variables used to determine $\eta_t$.\\

To impose these requirements, define the instrument vector $\mathbf{Z}_t
=
\begin{pmatrix}
1\\
\mathbf{q}_t
\end{pmatrix}$. The first element, equal to one, ensures that the cross-moment residuals are
approximately zero on average. The remaining elements are the state
variables themselves and ensure that the residuals are not systematically
related to those state variables. The corresponding GMM moment conditions are
\[
E\left[
\mathbf{Z}_t
\otimes
\mathbf{e}_t(\boldsymbol{\theta}_0)
\right]
=
\mathbf{0},
\]
where $\otimes$ denotes the Kronecker product. This expression simply
collects the conditions described above into a single vector. The parameters in $\boldsymbol{\theta}$ are therefore estimated so that,
across the sample, the observed cross-products are as close as possible to
the model-implied cross-moments and any remaining differences are not
systematically associated with the state variables. \\

For a sample of $N$ observations, define $\mathbf{g}_t(\boldsymbol{\theta})
=
\mathbf{Z}_t
\otimes
\mathbf{e}_t(\boldsymbol{\theta})$, 
and the sample-average moment vector $\overline{\mathbf{g}}_N(\boldsymbol{\theta})
=
\frac{1}{N}
\sum_{t=1}^{N}
\mathbf{g}_t(\boldsymbol{\theta})$. The dynamic dependence parameters are estimated as
\[
\widehat{\boldsymbol{\theta}}
=
\arg\min_{\boldsymbol{\theta}}
\;
\overline{\mathbf{g}}_N(\boldsymbol{\theta})^\top
W
\overline{\mathbf{g}}_N(\boldsymbol{\theta}),
\]
where $W$ is a positive definite weighting matrix. In the empirical
implementation considered, the identity matrix is used so that all
moment conditions receive equal weight.\\

Once
$\widehat{\boldsymbol{\theta}}=
(\widehat{\beta}_0,
\widehat{\boldsymbol{\beta}}^\top)^\top$
has been obtained, the fitted dependence parameter is $\widehat{\eta}_t
=
\tanh\left(
\widehat{\beta}_0
+
\mathbf{q}_t^\top
\widehat{\boldsymbol{\beta}}
\right)$, and the corresponding fitted latent cross-covariance matrix is $\widehat{\Delta}_t
=
\Delta(\widehat{\eta}_t)$. Hence, a single set of coefficients
$\widehat{\boldsymbol{\theta}}$ determines an entire sequence of dependence
matrices $\widehat{\Delta}_1,
\widehat{\Delta}_2,
\ldots,
\widehat{\Delta}_N$. The model therefore remains parsimonious: it does not estimate a separate
free dependence parameter for every observation. 

\subsection{Numerical evaluation of the mixed moments}
We present two approaches for computing the mixed moments. The first uses Monte Carlo simulation to approximate the Dirichlet average in Carlson's multiple R-function. The second uses a Laplace transform to express the moments as two-dimensional integrals, which we evaluate by deterministic quadrature.
\subsubsection{Monte Carlo evaluation of the multiple
$R$-function}

Although the mixed moments admit the closed-form representation in
terms of Carlson's multiple $R$-function, direct evaluation
of this function is not required. Following
\citet{JiangKadaneDickey1992}, the Dirichlet-average representation, as found in Lemma \ref{lem:carlson_multiple_R},
provides a simple Monte Carlo method for its numerical evaluation:\\

For a candidate value of $\boldsymbol\Delta$, let $M=K+L+KL$ and generate independently 
\[
G_m^{(s)}
\sim
\operatorname{Gamma}
\left(
\gamma_m(\boldsymbol\Delta),1
\right),
\qquad
m=1,\ldots,M,
\]
for $s=1,\ldots,S$. After defining $Z_m^{(s)}
=
\frac{
G_m^{(s)}
}{
\sum_{h=1}^{M}G_h^{(s)}
}$, it follows that $\mathbf Z^{(s)}
\sim
\operatorname{Dirichlet}
\left(
\boldsymbol\gamma(\boldsymbol\Delta)
\right)$.\\

Hence, for $\mathbf H
=
[\mathbf h_1,\ldots,\mathbf h_J]$
and exponent vector $\boldsymbol\nu
=
(\nu_1,\ldots,\nu_J)^\top$, 
the multiple $R$-function is approximated by
\[
\widehat{\mathcal R}_{S}
\left(
\boldsymbol\gamma;
\mathbf H;
\boldsymbol\nu
\right)
=
\frac{1}{S}
\sum_{s=1}^{S}
\prod_{j=1}^{J}
\left[
\mathbf h_j^\top
\mathbf Z^{(s)}
\right]^{\nu_j}.
\]

\subsubsection{Deterministic quadrature evaluation of the mixed moments}

Although the mixed moments have been expressed in terms of the multiple
$R$-function, direct numerical evaluation of the $R$-function is not required.
Instead, the theoretical mixed moments used in the second stage can be
evaluated deterministically using a Laplace-transform representation and
numerical quadrature.\\

For convenience, define the two normalising sums $S_X
=
\sum_{h=1}^{K}\rho_hU_h$ and $S_Y
=
\sum_{m=1}^{L}\kappa_mV_m$. Since $X_j
=
\frac{\rho_jU_j}{S_X}$ and $Y_\ell
=
\frac{\kappa_\ell V_\ell}{S_Y}$, a mixed moment of positive integer orders $r$ and $s$ can be written as $\mathbb{E}(X_j^rY_\ell^s)
=
\rho_j^r\kappa_\ell^s
\mathbb{E}
\left[
\frac{U_j^rV_\ell^s}
{S_X^rS_Y^s}
\right]$. The inverse powers of the two denominators can be represented using the
Gamma integral identity
\[
z^{-r}
=
\frac{1}{\Gamma(r)}
\int_{0}^{\infty}
u^{r-1}e^{-uz}\,du,
\qquad z>0.
\]
Applying this identity separately to $S_X^{-r}$ and $S_Y^{-s}$ gives
\[
\mathbb{E}(X_j^rY_\ell^s)
=
\frac{
\rho_j^r\kappa_\ell^s
}{
\Gamma(r)\Gamma(s)
}
\int_{0}^{\infty}
\int_{0}^{\infty}
u^{r-1}v^{s-1}
\mathbb{E}
\left[
U_j^rV_\ell^s
e^{-uS_X-vS_Y}
\right]
\,du\,dv.
\]
The expectation inside the integral can be evaluated analytically using the
latent Gamma construction. In particular, since $U_k
=
A_k+\sum_{l=1}^{L}D_{k\ell}$ and $
V_\ell
=
B_\ell+\sum_{k=1}^{K}D_{k\ell}$, the powers
$U_j^r$ and $V_\ell^s$ involve powers of sums of independent Gamma variables. These powers are therefore expanded so that each resulting term is
expressed as a product of powers of the Gamma variables
$A_k$, $B_\ell$ and $D_{k\ell}$. This is useful because these Gamma variables are independent and expectations involving
each individual Gamma variable can be evaluated in closed form using
\[
\mathbb{E}
\left(
G^q e^{-cG}
\right)
=
(\alpha)_q
(1+c)^{-(\alpha+q)},
\]
where $G\sim\operatorname{Gamma}(\alpha,1)$ with $(\alpha)_q
=
\frac{\Gamma(\alpha+q)}{\Gamma(\alpha)}$. \\

After this analytical step, only a two-dimensional numerical integral remains.
For numerical evaluation, the integration region is transformed to
the unit square using
\[
u=\frac{x}{1-x},
\qquad
v=\frac{y}{1-y},
\qquad
0<x<1,\quad 0<y<1,
\]
with Jacobian terms $du=\frac{dx}{(1-x)^2}$ and $dv=\frac{dy}{(1-y)^2}$.\\

Consequently, the required integral can be written over $[0,1]^2$ and
evaluated using product Gauss-Legendre quadrature. If
$x_g$ and $w_g$, $g=1,\ldots,Q$, denote the Gauss-Legendre nodes and
weights on $[0,1]$, then an integral of the form $\int_0^1\int_0^1 h(x,y)\,dx\,dy$ is approximated by
\[
\int_0^1\int_0^1 h(x,y)\,dx\,dy
\approx
\sum_{g=1}^{Q}
\sum_{h=1}^{Q}
w_gw_h
h(x_g,x_h).
\]

An important computational advantage of this representation is that the
dimension of the numerical integration does not increase with the number of
components in the compositions. Regardless of $K$, there are only two
normalising denominators, $S_X$ and $S_Y$, and hence only two integration
variables, $u$ and $v$. The theoretical mixed moments can therefore be
evaluated using two-dimensional quadrature even when the dimension of the
compositions increases.
% Unlike Monte Carlo evaluation, this procedure introduces no simulation noise:
% for a given value of $\omega$ and the marginal parameters, the same
% theoretical mixed moments are obtained at every evaluation. This is
% particularly useful during optimisation, since the objective function used
% to estimate $\omega$ is then smooth and reproducible. Numerical accuracy can
% be checked by increasing the quadrature order $Q$ until the calculated mixed
% moments remain effectively unchanged.
\paragraph{Numerical implementation for the time-varying case.}
For given marginal parameter estimates, the model-implied cross-moments
depend on the dynamic specification only through the value of $\eta_t$.
Calculating these moments by numerical quadrature separately for every
observation, and repeatedly for every candidate set of GMM parameters,
would be computationally expensive.\\

To avoid this, the theoretical moment vector $\mathbf{m}(\eta)$ is first
calculated over a dense grid of values between $-1$ and $1$. These
calculations are performed once before the GMM optimisation. During
estimation, each candidate parameter vector produces an observation-specific
value $\eta_t$. The corresponding theoretical moments are then obtained
from the pre-calculated grid by interpolation between the nearest values of
$\eta$.

\subsection{Out-of-sample model evaluation}

The out-of-sample performance of the coupled scaled Dirichlet distribution is evaluated using the energy score and variogram score, both of which are proper scoring rules for multivariate distributions as discussed in \citet{gneiting2007strictly,scheuerer2015variogram}. Proper scoring rules assess the entire predictive distribution rather than a single feature such as the mean or covariance, with lower scores indicating better predictive performance.\\

To ensure that the comparison reflects out-of-sample performance, the data are divided into a training sample and a holdout test sample. All model parameters are estimated using only the training sample. For the coupled scaled Dirichlet model, this entails first estimating the marginal parameters $\mathbf a$, $\boldsymbol{\rho}$, $\mathbf b$, and $\boldsymbol{\kappa}$, followed by estimation of the dependence parameter $\eta$ from the selected mixed moments. The corresponding parameters of the coupled ordinary Dirichlet model are estimated from the same training observations. Once estimation is complete, the parameters are held fixed and observations are simulated from each fitted joint distribution\footnote{Although a closed-form joint density for $(\mathbf X,\mathbf Y)$ is not required for the proposed estimation and evaluation procedure, the coupled model has an explicit latent-variable representation from which joint observations can be simulated directly. Conditional on the estimated parameters, independent shared Gamma variables
$D_{k\ell}\sim\operatorname{Gamma}(\delta_{k\ell},1)$
are generated for all $(k,\ell)$ such that $\delta_{k\ell}>0$. Independent private Gamma variables are then generated as $A_k\sim\operatorname{Gamma}\left(
a_k-\sum_{\ell=1}^{L}\delta_{k\ell},1
\right)$ for $k=1,\ldots,K$ 
and $B_\ell\sim\operatorname{Gamma}\left(
b_\ell-\sum_{k=1}^{K}\delta_{k\ell},1
\right)$ for $\ell=1,\ldots,L$, where the elements $\delta_{k\ell}$ are determined by the static dependence parameter $\eta$.
The latent Gamma variables are subsequently constructed as $U_k
=
A_k+\sum_{\ell=1}^{L}D_{k\ell}$ and $V_\ell
=
B_\ell+\sum_{k=1}^{K}D_{k\ell}$. Finally, the compositions are obtained by scaled normalisation, $X_k
=
\frac{\rho_kU_k}
{\sum_{h=1}^{K}\rho_hU_h}$ and $
Y_\ell
=
\frac{\kappa_\ell V_\ell}
{\sum_{m=1}^{L}\kappa_mV_m}$. Repeating this procedure yields Monte Carlo draws from the fitted joint distribution of $(\mathbf X,\mathbf Y)$ while preserving the scaled-Dirichlet marginal distributions, even though the corresponding joint density is not available in closed form.
}. These simulated observations are then used to evaluate the predictive distribution against each observation in the holdout test sample. Thus, the test observations play no role in parameter estimation.\\

Since the observations are compositional, the scoring rules are applied after transforming the simulated and observed compositions to Euclidean coordinates. In particular, each composition is transformed separately using the isometric log-ratio (ilr) transformation (\citet{aitchison1986statistical,egozcue2003isometric}). For $\mathbf X\in\mathcal S^{K-1}$ and $\mathbf Y\in\mathcal S^{L-1}$, define

$$
\mathbf z
=
\begin{pmatrix}
\operatorname{ilr}(\mathbf X)\\
\operatorname{ilr}(\mathbf Y)
\end{pmatrix}
\in\mathbb R^{K+L-2}.
$$

The same transformation and ilr basis are applied to both the observed test compositions and the observations simulated from the fitted model. This allows the scoring rules to be evaluated in an unconstrained Euclidean space while respecting the geometry of compositional data. The energy and variogram scores are therefore calculated using the transformed vector $\mathbf z$, rather than directly from the raw compositional components.\\

Let $\mathbf Z$ denote a random vector from a fitted joint distribution $F$, let $\mathbf Z'$ be an independent copy of $\mathbf Z$, and let $\mathbf z$ denote an observed test vector. The energy score is

$$
\operatorname{ES}(F,\mathbf z)
=
\operatorname{E}_{F}
\left\|
\mathbf Z-\mathbf z
\right\|
-
\frac{1}{2}
\operatorname{E}_{F}
\left\|
\mathbf Z-\mathbf Z'
\right\|,
$$

where $|\cdot|$ denotes the Euclidean norm (\citet{gneiting2007strictly}). The first term rewards concentration of the fitted distribution around the realised observation, while the second accounts for the dispersion of the predictive distribution. The energy score therefore provides an overall assessment of multivariate distributional fit.\\

The variogram score provides a complementary assessment that places greater emphasis on the dependence structure between components (\citet{scheuerer2015variogram}). For order $p>0$, it is defined as

$$
\operatorname{VS}_{p}(F,\mathbf z)
=
\sum_{j=1}^{d}
\sum_{k=1}^{d}
w_{jk}
\left[
|z_j-z_k|^{p}
-
\operatorname{E}_{F}
\left(
|Z_j-Z_k|^{p}
\right)
\right]^2,
$$

where $w_{jk}\geq0$ are weights and $d$ is the dimension of the joint observation. The variogram score compares pairwise differences between components of the observed vector with those implied by the fitted joint distribution and is therefore particularly informative when comparing alternative dependence structures. In the present application, $p=0.5$ is used.\\

For each model, the energy and variogram scores are calculated for every observation in the holdout sample and then averaged across the test observations. The preferred model is the one with the lower mean out-of-sample score. Using both measures is useful because the energy score provides a broad assessment of the joint predictive distribution, whereas the variogram score provides additional sensitivity to how well the dependence among its components is reproduced.

\section{Simulation Study}

A Monte Carlo study is conducted to evaluate the finite-sample performance of
the full two-stage estimator and, in particular, its ability to recover the
observable dependence structure implied by the proposed model.

\subsection{Static signed dependence model}
\subsubsection{Simulation design}

Data are generated from the coupled scaled-Dirichlet model with $K=3$ and
$L=4$. The marginal parameters are fixed at
$\mathbf{a}=(4,6,5)^\top, \boldsymbol{\rho}=(1.5,0.8,1)^\top, \mathbf{b}=(5,7,6,4)^\top$ and $
\boldsymbol{\kappa}=(0.8,1.2,1,0.9)^\top$. The first three components are treated as corresponding pairs. Thus, for
$\eta>0$, shared Gamma components connect $(X_1,Y_1)$, $(X_2,Y_2)$, and
$(X_3,Y_3)$, while $Y_4$ has no matched counterpart. For $\eta<0$, sharing
occurs across non-matched component pairs, with the crossed contribution
normalised by $\max(K,L)-1=3$.\\

The dependence parameter is considered under both crossed and matched
dependence where $\eta\in\{-0.95,-0.55,-0.15,\\ 0.15,0.55,0.95\}$, and the sample size is varied over $N\in\{250,500,1000\}$. For each combination of $\eta$ and $N$, $R = 1000$ independent samples are generated. In every replication, the marginal scaled-Dirichlet parameters are first
estimated separately by the method of moments as the first stage. Conditional on these estimates,
$\eta$ is then estimated from the $(K-1)(L-1)=6$ independent second-order
cross-moments using the identity-weighted minimum-distance criterion.

\subsubsection{Dependence recovery}

The simulation evaluates both recovery of the dependence parameter and recovery
of the resulting observable dependence structure. For each value of $\eta$, a
large sample generated from the true model is used to approximate the population
cross-correlation matrix: $\mathbf{C}_0=\operatorname{Corr}(\mathbf{X},\mathbf{Y})$. After estimating the full model in each Monte Carlo replication, a large sample
is generated from the fitted model to obtain the fitted cross-correlation matrix
$\widehat{\mathbf{C}}$. This comparison is important because $\eta$ controls dependence through the latent Gamma variables, while the dependence observed between
$\mathbf X$ and $\mathbf Y$ is additionally affected by the normalisation. Consequently, successful recovery of $\eta$ does not
automatically imply exact recovery of the observed cross-correlation matrix.\\

Correlation recovery is assessed using $\operatorname{RMSE}_{C}
=
\left[
\frac{1}{KL}
\sum_{k=1}^{K}\sum_{\ell=1}^{L}
\left(
\widehat C_{k\ell}-C_{0,k\ell}
\right)^2
\right]^{1/2}$ together with the correlation
between the vectorised fitted and population correlation matrices as $\operatorname{Corr}
\left(
\operatorname{vec}(\widehat{\mathbf C}),
\operatorname{vec}(\mathbf C_0)
\right)$. Estimation of $\eta$ is additionally evaluated
using its bias and RMSE and the frequency with which the correct dependence
branch is recovered.

\subsubsection{Simulation results}
\begin{figure}[H]
    \centering
    \begin{minipage}[t]{0.48\textwidth}
        \centering
        \includegraphics[width=\linewidth]{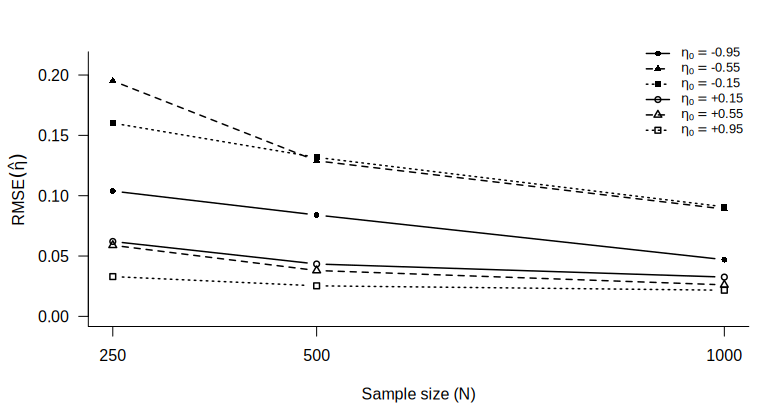}
        \caption{Root mean squared error of the estimated dependence
        parameter $\hat{\eta}$ across sample sizes and true parameter
        values $\eta_0$.}
        \label{fig:eta_rmse}
    \end{minipage}
    \hfill
    \begin{minipage}[t]{0.48\textwidth}
        \centering
        \includegraphics[width=\linewidth]{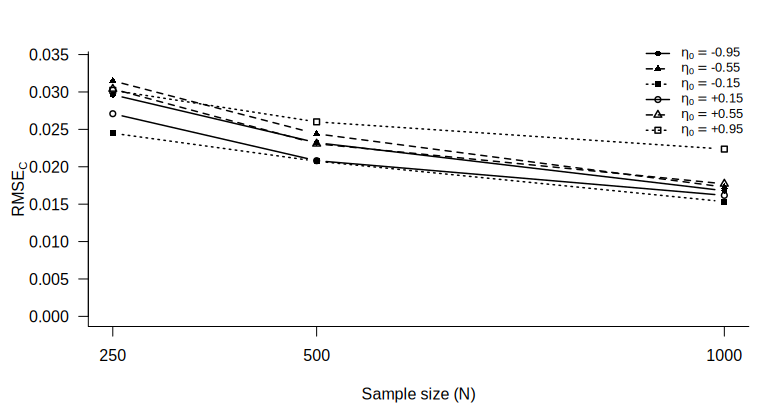}
        \caption{Root mean squared error of the estimated
        cross-correlations across sample sizes and true dependence
        parameter values $\eta_0$.}
        \label{fig:cross_corr_rmse}
    \end{minipage}
\end{figure}

Table~\ref{tab:mc_full} and Figures~\ref{fig:eta_rmse} and
\ref{fig:cross_corr_rmse} summarise the Monte Carlo results. Overall, the
two-stage estimator recovers $\eta$ accurately, with generally small bias and
declining RMSE as the sample size increases. Recovery is particularly strong
under moderate and strong dependence, with the correct matched or crossed
dependence regime identified in almost all replications.\\

Estimation is less precise under weak crossed dependence. For
$\eta_0=-0.15$, branch recovery increases from $0.74$ at $N=250$ to $0.83$ at
$N=1000$, while RMSE$_{\eta}$ declines from $0.160$ to $0.091$. This is
consistent with weaker identification of the dependence parameter close to
independence on the crossed branch.\\

Recovery of the observable dependence structure also improves with sample
size. RMSE$_C$ decreases across all parameter configurations, while the
correlation between the fitted and population cross-correlation matrices is
close to $1$ under moderate and strong dependence. These results indicate
that the proposed estimator recovers both the underlying static dependence
parameter and the cross-correlation structure generated by the model.

\begin{table}[H]
\centering
\small
\begin{tabular}{rrrrrrrr}
\hline
$\eta_0$ & $N$ & Mean $\widehat{\eta}$ & Bias &
RMSE$_{\eta}$ & Branch recovery &
RMSE$_{C}$ & Matrix corr. \\
\hline

-0.95 & 250  & -0.929 &  0.021 & 0.104 & 1.00 & 0.030 & 0.987 \\
-0.95 & 500  & -0.936 &  0.014 & 0.084 & 1.00 & 0.023 & 0.991 \\
-0.95 & 1000 & -0.965 & -0.015 & 0.047 & 1.00 & 0.017 & 0.995 \\
\hline
-0.55 & 250  & -0.552 & -0.002 & 0.195 & 0.99 & 0.031 & 0.980 \\
-0.55 & 500  & -0.562 & -0.012 & 0.129 & 1.00 & 0.024 & 0.986 \\
-0.55 & 1000 & -0.566 & -0.016 & 0.089 & 1.00 & 0.017 & 0.991 \\
\hline
-0.15 & 250  & -0.163 & -0.013 & 0.160 & 0.74 & 0.024 & 0.597 \\
-0.15 & 500  & -0.185 & -0.035 & 0.132 & 0.80 & 0.021 & 0.768 \\
-0.15 & 1000 & -0.145 &  0.005 & 0.091 & 0.83 & 0.015 & 0.755 \\
\hline
 0.15 & 250  &  0.153 &  0.003 & 0.062 & 0.95 & 0.027 & 0.990 \\
 0.15 & 500  &  0.152 &  0.002 & 0.043 & 1.00 & 0.021 & 0.995 \\
 0.15 & 1000 &  0.144 & -0.006 & 0.033 & 1.00 & 0.016 & 0.995 \\
\hline
 0.55 & 250  &  0.556 &  0.006 & 0.059 & 1.00 & 0.030 & 0.998 \\
 0.55 & 500  &  0.554 &  0.004 & 0.038 & 1.00 & 0.023 & 0.998 \\
 0.55 & 1000 &  0.551 &  0.001 & 0.026 & 1.00 & 0.018 & 0.999 \\
\hline
 0.95 & 250  &  0.954 &  0.004 & 0.033 & 1.00 & 0.030 & 0.998 \\
 0.95 & 500  &  0.948 & -0.002 & 0.025 & 1.00 & 0.026 & 0.999 \\
 0.95 & 1000 &  0.949 & -0.001 & 0.022 & 1.00 & 0.022 & 0.999 \\

\hline
\end{tabular}

\vspace{0.2cm}

\begin{minipage}{0.95\textwidth}
\footnotesize
\caption{Monte Carlo performance of the full two-stage static $\eta$ estimator. Results are based on $R=1000$ Monte Carlo replications for
each combination of the true dependence parameter $\eta_0$ and sample size
$N$. RMSE$_{\eta}$ denotes the root mean squared error of
$\widehat{\eta}$. Branch recovery is the proportion of replications in which
the matched or crossed dependence branch is correctly identified.
RMSE$_C$ is the root mean squared error between the fitted and population
cross-correlation matrices. Matrix corr. denotes the correlation between
their vectorised entries.}
\label{tab:mc_full}
\end{minipage}
\end{table}

\subsection{Time-varying signed dependence model}
\subsubsection{Simulation design}
\label{sec:dynamic_sim_design}

A second Monte Carlo experiment is conducted to evaluate the performance of the time-varying signed-dependence estimator. The purpose of
this experiment is to determine whether the proposed estimation procedure can
recover both the parameters governing the dependence dynamics and the resulting
observation-specific dependence structure.\\

The data-generating process uses $K=L=3$. The marginal scaled-Dirichlet
parameters are fixed at $\mathbf{a}=(4,6,5)^\top$, $\boldsymbol{\rho}=(1.5,0.8,1)^\top, \mathbf{b}=(5,7,6)^\top$ and $\boldsymbol{\kappa}=(0.8,1.2,1)^\top$. Two state variables are generated independently as stationary AR(1)
processes: $q_{1t}
=
\phi_1q_{1,t-1}
+
\sqrt{1-\phi_1^2}\,\varepsilon_{1t}$ 
and $q_{2t}
=
\phi_2q_{2,t-1}
+
\sqrt{1-\phi_2^2}\,\varepsilon_{2t}$
where $\varepsilon_{1t},\varepsilon_{2t}
\overset{\mathrm{iid}}{\sim}N(0,1)$, 
with $\phi_1=0.85$ and $\phi_2=0.60$.
The initial values are generated from $N(0,1)$.\\

For each observation, the signed dependence parameter is generated according
to $\eta_t
=
\tanh
\left(
\beta_0+\beta_1q_{1t}+\beta_2q_{2t}
\right)$, 
with $\beta_0=0,
\beta_1=-0.90$ and $\beta_2=0.60$. Since the state variables are centred
at zero and $\beta_0=0$, the design generates observations on both dependence
branches rather than restricting the simulation to either matched or crossed
dependence. The full two-stage estimator is applied in every Monte Carlo replication.
Thus, the marginal parameters are first estimated from the simulated
compositions rather than treated as known. Conditional on these estimates,
the coefficients $\boldsymbol{\theta}
=
(\beta_0,\beta_1,\beta_2)^\top$ are estimated using the procedure described in Section~\ref{sec:dynamic_estimation}.\\

For $K=L=3$, four independent second-order cross-moments are used: $E[X_1Y_1], E[X_1Y_2],
E[X_2Y_1]$ and $
E[X_2Y_2]$. Together with the instrument vector $\mathbf Z_t=(1,q_{1t},q_{2t})^\top$. This produces $4\times3=12$ moment conditions for the three dynamic
dependence parameters. Sample sizes $N\in\{250,500,1000\}$ are considered, with $R=1000$ independent Monte Carlo replications for each
sample size. Identity weighting is used in the GMM objective.

\subsubsection{Dynamic dependence recovery}
\label{sec:dynamic_sim_recovery}

The simulation evaluates recovery at three levels. First, recovery of the
parameters governing the dynamic dependence process is assessed using the
bias and root mean squared error (RMSE) of
$\widehat{\beta}_0$, $\widehat{\beta}_1$ and $\widehat{\beta}_2$.\\

Second, the simulation assesses recovery of the complete dependence path.
For each replication, the RMSE between the fitted and true values of
$\eta_t$ is calculated as $\operatorname{RMSE}_{\eta}
=
\left[
\frac{1}{N}
\sum_{t=1}^{N}
(\widehat{\eta}_t-\eta_t)^2
\right]^{1/2}$. The correlation between the two paths, $\operatorname{Corr}
(\widehat{\boldsymbol{\eta}},\boldsymbol{\eta})$, is also reported. \\

Because the sign of $\eta_t$ determines the latent sharing structure, branch
recovery is assessed separately. Branch accuracy is defined as $\frac{1}{N}
\sum_{t=1}^{N}
\mathbf{1}
\left\{
\operatorname{sign}(\widehat{\eta}_t)
=
\operatorname{sign}(\eta_t)
\right\}$. This quantity measures the proportion of observations for which the model
correctly identifies the matched or crossed dependence branch.\\

Third, recovery of the observable dependence structure is assessed. Let $\mathbf{C}_t
=
\operatorname{Corr}
(\mathbf X_t,\mathbf Y_t\mid\mathbf q_t)$
denote the $K\times L$ population cross-correlation matrix implied by the
true value $\eta_t$, and let $\widehat {\mathbf {C}}_t$ denote the corresponding matrix
implied by the fitted model. Correlation recovery is measured using $\operatorname{RMSE}_{C}
=
\left[
\frac{1}{NKL}
\sum_{t=1}^{N}
\sum_{k=1}^{K}
\sum_{\ell=1}^{L}
\left(
\widehat C_{t,k\ell}-C_{t,k\ell}
\right)^2
\right]^{1/2}$. In addition, the correlation between the true and fitted
cross-correlation paths is calculated as $\operatorname{Corr}
\left[
\operatorname{vec}
(\mathbf C_1,\ldots,\mathbf C_N),
\operatorname{vec}
(\widehat {\mathbf C}_1,\ldots,\widehat{\mathbf C}_N)
\right]$.

\subsubsection{Simulation results}
Table~\ref{tab:dynamic_mc} shows systematic improvement in both parameter and
dependence-path recovery as the sample size increases. Bias and RMSE decline
for all three dynamic coefficients, although the slope coefficients are
estimated less precisely than the intercept. This improvement carries through
to the dependence-path process: the RMSE of $\widehat{\eta}_t$ decreases
from $0.271$ at $N=250$ to $0.156$ at $N=1000$, while the correlation between
the true and estimated dependence paths rises from $0.941$ to $0.982$.
Correct classification of the matched versus crossed regime similarly
increases from $90.2\%$ to $94.7\%$.\\

Recovery of the observable dependence structure improves concurrently.
Cross-correlation RMSE declines from $0.124$ to $0.074$, while agreement
between the true and fitted correlation paths increases from $0.932$ to
$0.979$. The results therefore indicate that the dynamic specification
recovers not only the coefficients governing $\eta_t$, but also the evolving
latent and observable dependence structures implied by those coefficients.

\begin{table}[H]
\centering
\small

\textbf{Panel A: Dynamic-parameter recovery}

\vspace{0.1cm}

\begin{tabular}{rllrrr}
\hline
$N$ & Parameter & True & Mean estimate & Bias & RMSE \\
\hline
250  & $\beta_0$              &  0.00 & -0.054 & -0.054 & 0.254 \\
250  & $\beta_{\mathrm{stress}}$    & -0.90 & -1.138 & -0.238 & 0.655 \\
250  & $\beta_{\mathrm{liquidity}}$ &  0.60 &  0.756 &  0.156 & 0.511 \\
\hline
500  & $\beta_0$              &  0.00 & -0.033 & -0.033 & 0.173 \\
500  & $\beta_{\mathrm{stress}}$    & -0.90 & -1.032 & -0.132 & 0.486 \\
500  & $\beta_{\mathrm{liquidity}}$ &  0.60 &  0.680 &  0.080 & 0.369 \\
\hline
1000 & $\beta_0$              &  0.00 & -0.025 & -0.025 & 0.128 \\
1000 & $\beta_{\mathrm{stress}}$    & -0.90 & -0.996 & -0.096 & 0.367 \\
1000 & $\beta_{\mathrm{liquidity}}$ &  0.60 &  0.662 &  0.062 & 0.278 \\
\hline
\end{tabular}

\vspace{0.45cm}

\textbf{Panel B: Time-varying dependence recovery}

\vspace{0.1cm}

\begin{tabular}{rrrrrr}
\hline
$N$
& RMSE$_{\eta}$
& Corr.$(\eta_t,\widehat{\eta}_t)$
& Branch accuracy
& RMSE$_C$
& Corr. path \\
\hline
250  & 0.271 & 0.941 & 0.902 & 0.124 & 0.932 \\
500  & 0.205 & 0.968 & 0.928 & 0.095 & 0.962 \\
1000 & 0.156 & 0.982 & 0.947 & 0.074 & 0.979 \\
\hline
\end{tabular}

\vspace{0.15cm}

\begin{minipage}{0.94\textwidth}
\footnotesize
\caption{Monte Carlo performance of the time-varying (dynamic) signed $\eta$ model. Panel A reports recovery of the coefficients governing the
time-varying dependence parameter. Panel B evaluates recovery of the resulting
dependence process. RMSE$_{\eta}$ is the root mean squared error between the
true and estimated paths $\eta_t$ and $\widehat{\eta}_t$.
Corr.$(\eta_t,\widehat{\eta}_t)$ measures their association.
Branch accuracy is the proportion of observations for which the estimated
sign of $\eta_t$ correctly identifies the matched or crossed sharing regime.
RMSE$_C$ measures error in the implied time-varying cross-correlation
structure, while Corr. path measures agreement between the true and fitted
cross-correlation paths.}
\label{tab:dynamic_mc}
\end{minipage}
\end{table}

\section{Application}
To illustrate the proposed model, we consider daily data for four major U.S.
equity-index futures contracts: the E-mini S\&P 500 (ES), E-mini Nasdaq-100
(NQ), E-mini Dow Jones (YM), and E-mini Russell 2000 (RTY). Daily price and
trading-volume data are obtained from Yahoo Finance for the period January
2018 to September 2026, yielding 2,178 observations.\\

For each trading day, two four-part compositions are constructed. The first,
$\mathbf X_t$, represents the share of total trading volume attributable to
each $k^{th}$ futures contract, $X_{kt}
=
\frac{\mathrm{Volume}_{kt}}
{\sum_{j=1}^{4}\mathrm{Volume}_{jt}}$. The second, $\mathbf Y_t$, measures the relative contribution of each contract
to total intraday price variation. Intraday variation is measured using the
Parkinson range, $R_{kt}
=
\left[
\log\left(
\frac{H_{kt}}{L_{kt}}
\right)
\right]^2$ and converted to a composition as $Y_{kt}
=
\frac{R_{kt}}
{\sum_{j=1}^{4}R_{jt}}$. Thus, both $\mathbf X_t$ and $\mathbf Y_t$ lie on the simplex.\\

The application asks whether the proposed coupled scaled-Dirichlet
model can represent the dependence between the cross-sectional
allocation of trading activity and the cross-sectional allocation of intraday
price variation. 

\subsection{Static dependence results}

The final sample contains 2,178 paired daily observations from 2 January 2018
to 11 September 2026. For the descriptive analysis, the model is estimated using the full sample. Table~\ref{tab:empirical_parameters} reports the resulting marginal parameter
estimates. The dependence estimate is $\widehat{\eta}=0.551$, placing the fitted model on the matched-dependence branch. Thus, the empirical
fit favours shared latent Gamma components between corresponding futures
contracts rather than crossed sharing. \\

\begin{table}[H]
\centering
\small
\begin{tabular}{lrrrr}
\hline
& $\widehat a_k$ & $\widehat\rho_k$
& $\widehat b_k$ & $\widehat\kappa_k$ \\
\hline
ES  & 126.979 & 1.071 & 18.643 & 0.185 \\
NQ  &  31.236 & 1.518 &  4.090 & 1.709 \\
YM  &  13.156 & 1.072 &  4.911 & 0.717 \\
RTY &  15.790 & 1.000 &  8.086 & 1.000 \\
\hline
\end{tabular}

\vspace{0.15cm}

\begin{minipage}{0.92\textwidth}
\footnotesize
\caption{Full-sample parameter estimates for the coupled scaled-Dirichlet model. The parameters $(a_k,\rho_k)$ describe the marginal
scaled-Dirichlet distribution of the trading-volume composition, while
$(b_k,\kappa_k)$ describe the marginal distribution of the Parkinson-range
composition. The final relative scale parameter in each margin is fixed to one
for identification.}
\label{tab:empirical_parameters}
\end{minipage}
\end{table}

\begin{figure}[H]
    \centering
    \includegraphics[width=0.75\textwidth]{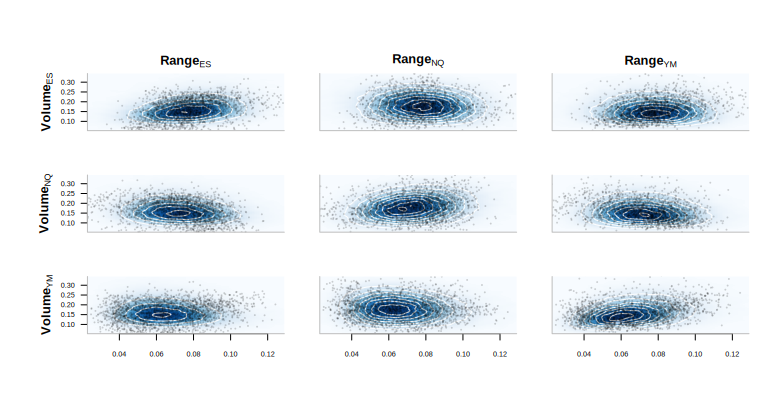}
\caption{Bivariate empirical observations and model-implied density contours of the trading-volume and Parkinson-range
compositions. Points represent the observed daily compositional pairs, while
the contour lines are obtained from the fitted coupled scaled-Dirichlet model.
Only ES, NQ and YM are displayed because a four-part composition has three
free coordinates; the RTY share can be determined by the simplex constraint, so is excluded here.}
    \label{fig:equity4-contours}
\end{figure}

Figure~\ref{fig:equity4-contours} provides a complementary visual assessment of the
fitted joint distribution. Across the nine free-coordinate pairs, the
model-implied contours generally place the highest probability mass in the
regions of greatest empirical concentration. In this sense, the fitted model
passes a basic visual plausibility check: despite the restrictions imposed by
the single dependence parameter $\eta$, it reproduces the principal orientation and concentration of the observed bivariate distributions
reasonably well. 

% Because the second estimation stage uses nine independent cross-moments to
% estimate a single parameter, the model is overidentified and is not expected
% to reproduce every empirical cross moment exactly. The fitted moment profiles, as shown in Figure~\ref{fig:moment_eta}, 
% therefore provide a direct indication of which aspects of the observed
% dependence structure can be accommodated by the one-parameter specification. From  Figure~\ref{fig:moment_eta}, most empirical moments lie close to
% their model-implied values at $\widehat{\eta} = 0.551$, although some discrepancies
% remain. These discrepancies are reflected in the empirical-versus-fitted
% cross-correlation comparison discussed below.

\begin{figure}[H]
    \centering
    \includegraphics[width=0.75\textwidth]{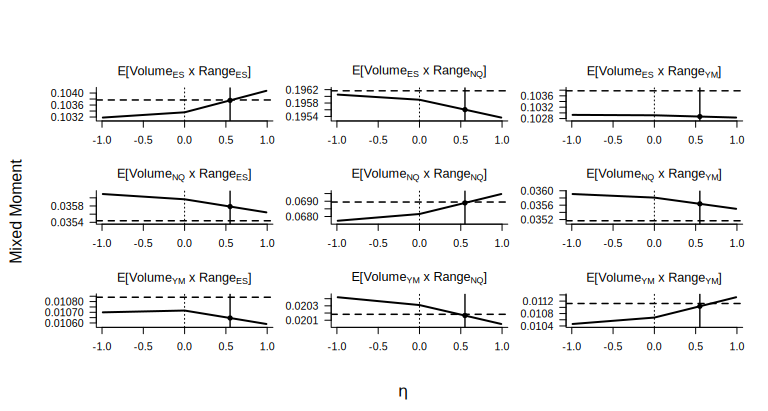}
\caption{Moment profiles for the nine independent mixed moments
used to estimate the dependence parameter $\eta$. In each panel, the
solid curve shows the model-implied moment $E[X_kY_\ell]$ as a function of
$\eta$, conditional on the estimated marginal parameters. The horizontal
dashed line denotes the corresponding empirical mixed moment, the vertical
dotted line at $\eta=0$ marks independence, and the vertical solid line marks
the estimated dependence parameter $\widehat{\eta}=0.551$. The solid point
indicates the model-implied moment evaluated at $\widehat{\eta}$. Differences
between the horizontal dashed lines and the fitted points illustrate the
extent to which a single common dependence parameter can jointly accommodate
the nine moment conditions.}
    \label{fig:equity4-moment-profiles}
\end{figure}

Figure~\ref{fig:equity4-moment-profiles} provides a diagnostic interpretation of the
dependence fit. Each panel plots one of the nine independent mixed moments,
$E[X_kY_\ell]$, as a function of the dependence parameter $\eta$,
holding the estimated marginal parameters fixed. The horizontal dashed line
represents the corresponding empirical mixed moment, while the vertical line
indicates the estimated value $\widehat{\eta}=0.551$. The point on each curve
therefore gives the model-implied value of that moment at the common estimated
dependence parameter.\\

The figure highlights an important restriction of the proposed parsimonious
dependence structure. A single value of $\eta$ must jointly accommodate
all nine independent cross-moment conditions. Consequently, although some
moments are reproduced very closely at $\widehat{\eta}$, others would require
substantially different values of $\eta$ to match their empirical counterparts.
For some moment conditions, the empirical value also lies close to the edge of,
or outside, the range generated by the one-parameter moment curve. The fitted
value of $\eta$ should therefore be interpreted as the minimum-distance
compromise across the full set of dependence conditions rather than as a value
that reproduces each cross moment individually.\\

\begin{figure}[H]
    \centering
    \includegraphics[width=0.75\textwidth]{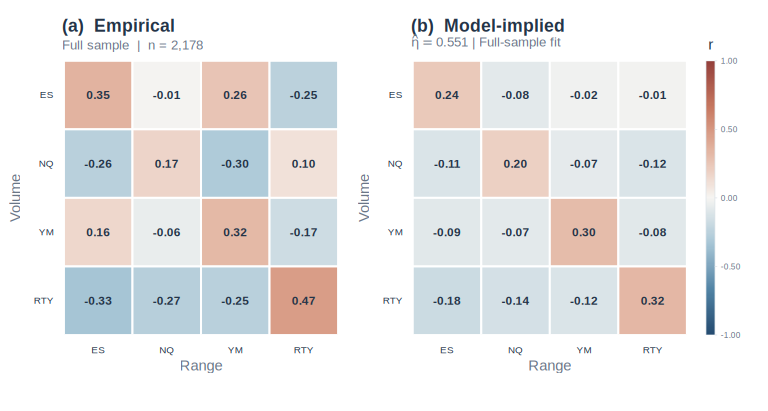}
\caption{Empirical and model-implied cross-correlation matrices for the
trading-volume and Parkinson-range compositions. Panel (a) reports the sample
cross-correlations computed from the full set of 2,178 observations, while
Panel (b) reports the corresponding correlations implied by the fitted coupled
scaled-Dirichlet model using the full-sample estimate
$\widehat{\eta}=0.551$. Rows correspond to the trading-volume compositions and
columns to the Parkinson-range compositions of ES, NQ, YM and RTY.}
    \label{fig:equity4-correlations}
\end{figure}

This restriction also explains the discrepancies between the empirical and
model-implied cross-correlation matrices in
Figure~\ref{fig:equity4-correlations}. For any pair $(X_k,Y_\ell)$, $\operatorname{Corr}(X_k,Y_\ell)
=
\frac{
E[X_kY_\ell]-E[X_k]E[Y_\ell]
}{
\sqrt{\operatorname{Var}(X_k)\operatorname{Var}(Y_\ell)}
}$. Once the marginal distributions have been estimated, the marginal means and
variances in this expression are largely determined by the Stage~1 parameter
estimates. The ability of the model to reproduce the observed
cross-correlations therefore depends importantly on its ability to reproduce
the mixed moments $E[X_kY_\ell]$. Because a single common value of $\eta$
must accommodate all nine independent mixed moments simultaneously,
discrepancies in the fitted moments translate directly into discrepancies in
the implied cross-correlations.\\

This relationship is particularly apparent for several coordinate pairs.
For $\mathrm{Volume}_{ES}$ and $\mathrm{Range}_{YM}$, the empirical mixed
moment is $0.10378$, compared with a model-implied value of $0.10287$ at
$\widehat{\eta}$ (seen in the top right panel in Figure \ref{fig:equity4-moment-profiles}). Although this difference is small in absolute terms, it is
sufficient relative to the product of the marginal means to produce a significant
difference in correlation: the empirical correlation is $0.26$, whereas the
model implies approximately $-0.02$ (seen in Figure \ref{fig:equity4-correlations}). A similar pattern occurs for
$\mathrm{Volume}_{YM}$ and $\mathrm{Range}_{ES}$, for which the empirical
and fitted mixed moments are $0.01084$ and $0.01065$, respectively; the
corresponding correlations are $0.16$ and $-0.09$. For
$\mathrm{Volume}_{NQ}$ and $\mathrm{Range}_{YM}$, the fitted mixed moment
($0.03564$) exceeds its empirical counterpart ($0.03517$), and the model
substantially understates the magnitude of the negative empirical correlation,
producing $-0.07$ compared with $-0.30$ in the data.\\

These cases illustrate why apparently modest deviations in the mixed moments
can generate much larger discrepancies in the resulting correlations. In
particular, when the empirical covariance is close to zero, a relatively
small difference in $E[X_kY_\ell]$ can move the covariance across zero and
therefore reverse the implied correlation sign. The discrepancies in
Figure~\ref{fig:equity4-correlations} are thus consistent with the moment profiles in
Figure~\ref{fig:equity4-moment-profiles}: they reflect the restriction imposed by using a
single dependence parameter to represent a set of pairwise
dependence relationships.\\

\subsubsection{Model comparison with ordinary coupled Dirichlet}
To assess whether the additional
flexibility of the scaled-Dirichlet translates into improved
distributional performance, the model is compared with a coupled ordinary
Dirichlet distribution ($\rho_k, \kappa_k = 1$ for $k = 1, \ldots, 4$), with the same $\eta$
dependence structure as before. For out-of-sample model comparison, both models are estimated on the first 80\% of the observations ($N=1,742$) and the remaining 436
observations are retained as a test sample. The training period
ends on 5 December 2024, with the test period beginning on 6 December 2024. \\

The fitted dependence parameter is positive
under both specifications, with $\widehat{\eta}_{\mathrm{SD}}=0.749$ and $\widehat{\eta}_{\mathrm{D}}=0.880$, 
so that both models select matched dependence. Table~\ref{tab:score_comparison} reports the held-out Energy and Variogram
scores. The scaled model obtains a slightly lower mean Energy Score
($0.71095$ versus $0.71136$), whereas the ordinary Dirichlet obtains a
slightly lower mean Variogram Score ($0.08791$ versus $0.08833$). The mean
scaled-minus-ordinary score differences are therefore $-0.00041$ for the
Energy Score and $0.00042$ for the Variogram Score. In both cases, the
moving-block bootstrap\footnote{To account for serial dependence in the held-out daily score differences, a moving-block bootstrap was applied to the sequence of paired scaled-minus-ordinary score differences. Overlapping blocks of length 21, chosen as $\operatorname{round}(\sqrt{436})$, were resampled with replacement to construct 2,000 bootstrap series of length 436. The mean score difference was computed for each bootstrap sample, and the 2.5th and 97.5th percentiles of the resulting distribution were used to form the 95\% confidence interval.} confidence interval contains zero. Consequently, the
held-out results provide no clear evidence that either model
dominates the other in predictive distributional performance.

\begin{table}[H]
\centering
\small

\begin{tabular}{lrrrrr}
\hline
Model
& $\widehat{\eta}$
& Mean ES
& Median ES
& Mean VS
& Median VS \\
\hline
Scaled Dirichlet
& 0.749
& 0.71095
& 0.60094
& 0.08833
& 0.07286 \\

Ordinary Dirichlet
& 0.880
& 0.71136
& 0.60562
& 0.08791
& 0.07321 \\
\hline
\end{tabular}

\vspace{0.35cm}

\begin{tabular}{lrrrr}
\hline
Score
& Mean difference
& 95\% CI
& Fraction scaled better \\
\hline
Energy
& -0.00041
& $[-0.00472,\;0.00335]$
& 0.569 \\

Variogram
& 0.00042
& $[-0.00016,\;0.00091]$
& 0.472 \\
\hline
\end{tabular}

\vspace{0.15cm}

\footnotesize
\caption{Held-out comparison of the coupled scaled-Dirichlet and ordinary
Dirichlet specifications. Lower Energy and Variogram scores indicate better
predictive performance. Score differences are calculated as scaled minus
ordinary, so negative values favour the scaled-Dirichlet model. The
``Fraction scaled better'' measure denotes the proportion of held-out days on
which the scaled model achieves a lower score than the ordinary model. The
reported 95\% confidence intervals for the mean score differences are based on
a moving-block bootstrap.}
\label{tab:score_comparison}
\end{table}

\subsection{Time-varying dependence results}
For the empirical application, the state vector contains two variables
intended to capture changes in financial-market conditions. The dynamic
dependence parameter is specified as
\[
\eta_t
=
\tanh
\left[
\beta_0
+
\beta_{\mathrm{VIX}}q_{\mathrm{VIX},t}
+
\beta_{10Y}q_{10Y,t}
\right],
\]
where $q_{\mathrm{VIX},t}$ is the standardised logarithm of the VIX index and
$q_{10Y,t}$ is the standardised daily change in the U.S. 10-year Treasury
yield, measured in basis points. Both variables are standardised using the
mean and standard deviation from the training sample only.\footnote{The static specification is estimated using all observations for which the
four futures series are jointly available. The time-varying specification
additionally requires observations of the VIX index and the
change in the 10-year U.S. Treasury yield. The futures data were downloaded separately for the time-varying analysis,
yielding 2,067 observations through 15 September 2026 after alignment
with the state variables. The static model is therefore retained on its larger available
sample when reported independently. For the direct out-of-sample comparison
between the static and time-varying specifications, however, both models are
estimated and evaluated using the same restricted sample and identical
training and test periods.}\\

Table~\ref{tab:dynamic_dependence_results} reports the estimated parameters
of the time-varying dependence specification. The VIX coefficient is positive
and substantially larger than the coefficient on changes in the 10-year
Treasury yield. Since both state variables are standardised, the coefficient
magnitudes are directly comparable. The estimates
therefore indicate that variation in the fitted dependence state is associated
primarily with changes in the VIX, while daily movements in the 10-year yield
make a comparatively small contribution.\\

The estimated dependence parameter varies substantially over the sample. In
the training period, $\widehat{\eta}_t$ ranges from $-0.979$ to approximately
$1$, with 39.1\% of observations assigned to the crossed branch and 60.9\%
to the matched branch. Thus, the static dependence structure would not
describe the fitted model: both the strength and the form of latent sharing
change across observations. The descriptive full-sample estimates are similar
to those obtained from the training sample, particularly for
$\beta_{\mathrm{VIX}}$ and $\beta_{10Y}$.

\begin{table}[H]
\centering
\begin{tabular}{lcc}
\hline
& \textbf{Training sample} & \textbf{Full sample} \\
\hline
\multicolumn{3}{l}{\textit{Dynamic dependence parameters}} \\[2pt]
$\widehat{\beta}_0$
    & 0.5209 & 0.5842 \\
$\widehat{\beta}_{\mathrm{VIX}}$
    & 1.2540 & 1.2397 \\
$\widehat{\beta}_{10Y}$
    & 0.0376 & 0.0372 \\[4pt]

\multicolumn{3}{l}{\textit{Fitted dependence path}} \\[2pt]
Minimum $\widehat{\eta}_t$
    & -0.9792 & -0.9749 \\
Maximum $\widehat{\eta}_t$
    & 0.9999 & 0.9999 \\
Crossed branch: $\widehat{\eta}_t<0$
    & 647 (39.14\%) & 718 (34.74\%) \\
Matched branch: $\widehat{\eta}_t\geq0$
    & 1,006 (60.86\%) & 1,349 (65.26\%) \\[4pt]
\hline
\end{tabular}

\begin{minipage}{0.92\textwidth}
\footnotesize
\caption{Time-varying signed-dependence estimates. The dynamic dependence parameter is
$\eta_t=\tanh(\beta_0+\beta_{\mathrm{VIX}}q_{\mathrm{VIX},t}
+\beta_{10Y}q_{10Y,t})$.
The state variables are the standardised logarithm of the VIX index and the
standardised daily change in the U.S. 10-year Treasury yield, respectively.
Standardisation uses the mean and standard deviation of the training sample.
Negative values of $\eta_t$ correspond to crossed sharing, while non-negative
values correspond to matched sharing. The training sample covers
3 January 2018 to 6 December 2024 (1,653 observations). The full-sample estimates are reported
descriptively using all observations through 15 September 2026 (2,067 observations).}
\label{tab:dynamic_dependence_results}
\end{minipage}
\end{table}

\begin{figure}[H]
\centering
\begin{subfigure}[t]{0.49\textwidth}
    \centering
    \includegraphics[width=\textwidth]{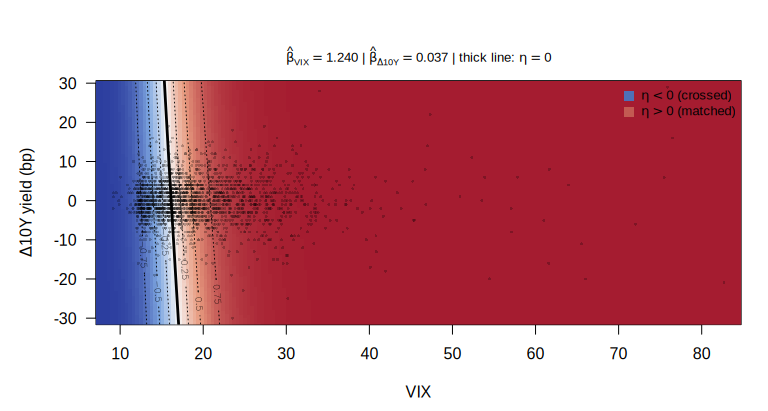}
    \caption{Contours of fitted dependence surface across VIX and 10-year Treasury yield-change states. The thick contour denotes $\widehat{\eta}_t=0$, separating crossed and matched dependence.}
    \label{fig:eta_state_space}
\end{subfigure}
\hfill
\begin{subfigure}[t]{0.49\textwidth}
    \centering
    \includegraphics[width=\textwidth]{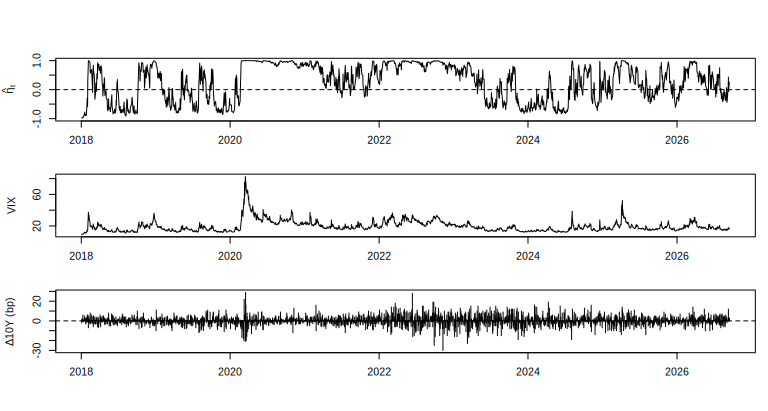}
    \caption{Time variation in the fitted dependence parameter, VIX, and the daily change in the 10-year Treasury yield.}
    \label{fig:eta_time_series}
\end{subfigure}

\caption{State-dependent variation in the fitted signed dependence parameter. For interpretability, the VIX and 10-year yield changes are shown in their original units, although
their standardised values enter the dynamic dependence specification.}
\label{fig:dynamic_dependence}
\end{figure}

Figure~\ref{fig:dynamic_dependence} illustrates how the estimated dependence
parameter varies with the market-state variables and through time.
Panel~\ref{fig:eta_state_space} shows that the fitted dependence surface is
driven primarily by the VIX. The nearly vertical contour lines indicate that
changes in the 10-year Treasury yield have comparatively little influence on
$\widehat{\eta}_t$, consistent with the substantially larger estimated
coefficient on the VIX reported in Table~\ref{tab:dynamic_dependence_results}.
Lower VIX states are associated predominantly with negative values of
$\widehat{\eta}_t$ and hence crossed sharing, whereas higher VIX states are
associated with positive values and matched sharing.\\

Panel~\ref{fig:eta_time_series} shows that the fitted dependence structure is
not constant over the sample. The estimated $\widehat{\eta}_t$ moves between
the crossed and matched branches and exhibits substantial variation in
magnitude. Periods of elevated VIX generally coincide with positive and often
relatively large values of $\widehat{\eta}_t$, whereas lower-volatility
periods are more frequently associated with crossed dependence. In contrast,
the relationship with daily changes in the 10-year Treasury yield is visually
much weaker.

\begin{figure}[H]
    \centering
    \includegraphics[width=0.75\textwidth]{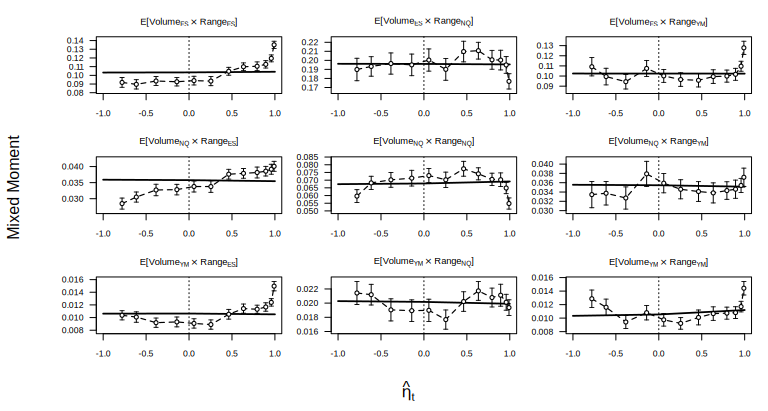}
    \caption{Conditional mixed-moment diagnostic for the time-varying signed-dependence model. The solid line in each panel represents the model-implied mixed moment as a function of $\eta$, while the open circles represent empirical mixed moments calculated within bins of the fitted $\widehat{\eta}_t$. Dashed lines connect the empirical bin means, and the vertical bars indicate approximate 95\% confidence intervals. The vertical dotted line at $\eta=0$ separates the crossed and matched dependence branches.}
    \label{fig:dynamic_moment_diagnostic}
\end{figure}

Figure~\ref{fig:dynamic_moment_diagnostic} evaluates whether the
time-varying signed-dependence specification reproduces the conditional
mixed-moment structure observed in the data. For each of the nine cross-moments, the solid curve represents the theoretical
mixed moment $E_{\eta}[X_kY_\ell]$ implied by the model as $\eta$ varies.
The empirical observations are grouped according to similar values of the
fitted dependence parameter $\widehat{\eta}_t$, and the open circles report
the average observed cross-product within each bin.\\

The correspondence between the empirical and model-implied moments varies
considerably across component pairs. For some relationships, the theoretical
curve captures the broad direction of the empirical variation. For others,
however, the empirical conditional moments display substantially greater
variation, non-monotonic behaviour, or persistent deviations from the
model-implied curve. In particular, several empirical moments remain above or
below the theoretical relationship over extended regions of the fitted
dependence range. Thus, allowing $\eta$ to vary with market conditions
improves the flexibility of the dependence specification, but a single
time-varying scalar parameter does not reproduce all features of the
conditional cross-moment structure.\\

\begin{figure}[H]
    \centering
    \includegraphics[width=0.75\textwidth]{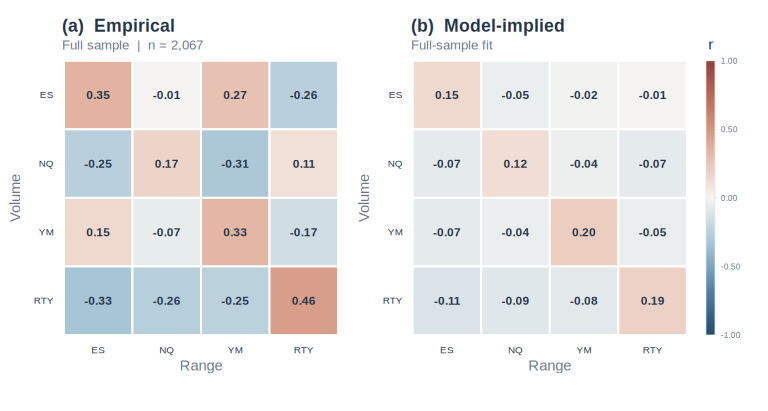}
    \caption{Empirical and model-implied cross-correlations between the volume and range compositions. Panel (a) reports the empirical full-sample correlations, while Panel (b) reports the correlations implied by the fitted time-varying signed-dependence model.}
    \label{fig:dynamic_correlation_heatmap}
\end{figure}

The discrepancies in the cross-correlation matrices in
Figure~\ref{fig:dynamic_correlation_heatmap} are consistent with the
mixed-moment diagnostic in Figure~\ref{fig:dynamic_moment_diagnostic}.
Although the fitted model captures some features of the empirical dependence structure, several model-implied correlations are weaker in magnitude and some have the wrong sign relative to the empirical correlations. This result follows naturally from the moment restrictions underlying the
model. For any component pair, $\operatorname{Cov}(X_k,Y_\ell)
=
E[X_kY_\ell]
-
E[X_k]E[Y_\ell]$, so that the model-implied cross-covariances, and consequently the
cross-correlations, depend directly on the corresponding mixed moments.
Figure~\ref{fig:dynamic_moment_diagnostic} shows that the empirical
conditional mixed moments do not lie uniformly close to the theoretical
moment curves. For some component pairs, the empirical moments vary more
strongly with $\widehat{\eta}_t$ than permitted by the model, while for
others they exhibit patterns that cannot be reproduced by changing
$\eta_t$ alone.\\

It should also be emphasised that the GMM estimator does not require each
empirical binned moment to coincide exactly with the theoretical curve in Figure \ref{fig:dynamic_moment_diagnostic}.
The dynamic specification contains three dependence parameters and is
estimated from 27 moment conditions. The estimator therefore selects the
parameter vector that minimises the joint moment discrepancies rather than
matching every conditional cross-moment separately. 

\subsubsection{Out-of-sample comparison of static and time-varying dependence}
Table~\ref{tab:static_dynamic_scores} compares the held-out distributional
performance of the static and time-varying signed-dependence specifications.
The dynamic model produces lower mean and median Energy and Variogram Scores
than the static model. The mean improvements are modest in magnitude, but
the paired score differences are negative for 71.0\% of held-out
observations under the Energy Score and 74.6\% under the Variogram Score.\\

The moving-block bootstrap confidence intervals for the mean score
differences lie entirely below zero for both scoring rules. Thus, allowing
$\eta_t$ to vary with the observed market-state variables yields a small but
systematic improvement in held-out distributional fit relative to the
constant-$\eta$ specification.\\

The distinction between the two specifications remains relevant
in the test sample: the static model estimates a single matched dependence
parameter of $\widehat{\eta}=0.748$, whereas the fitted dynamic parameter
ranges from $-0.604$ to approximately $1$, with 27.1\% of test observations
classified on the crossed branch.

\begin{table}[H]
\centering
\small

\begin{tabular}{lrrrr}
\hline
Model
& Mean ES
& Median ES
& Mean VS
& Median VS \\
\hline
Static $\eta$
& 0.71439
& 0.60904
& 0.08928
& 0.07276 \\

Dynamic $\eta_t$
& 0.71201
& 0.60663
& 0.08883
& 0.07267 \\
\hline
\end{tabular}

\vspace{0.35cm}

\begin{tabular}{lrrr}
\hline
Score
& Mean difference
& 95\% CI
& Fraction dynamic better \\
\hline
Energy
& -0.002385
& $[-0.002977,\;-0.001702]$
& 0.710 \\

Variogram
& -0.000443
& $[-0.000573,\;-0.000324]$
& 0.746 \\
\hline
\end{tabular}

\vspace{0.15cm}

\footnotesize
\caption{Held-out comparison of the static and time-varying signed-dependence
specifications. Lower Energy and Variogram scores indicate better predictive
performance. Score differences are calculated as dynamic minus static, so
negative values favour the time-varying model. The ``Fraction dynamic better''
measure denotes the proportion of held-out days on which the dynamic model
achieves a lower score than the static model. The reported 95\% confidence
intervals for the mean score differences are based on a moving-block
bootstrap.}
\label{tab:static_dynamic_scores}
\end{table}

\begin{figure}[H]
    \centering
    \includegraphics[width=0.75\textwidth]{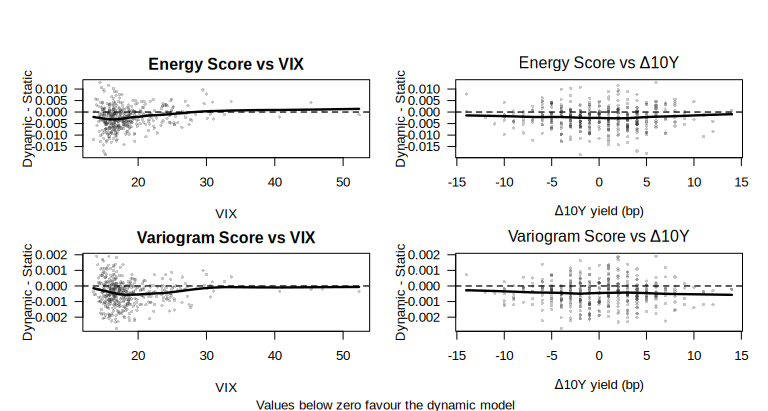}
    \caption{Held-out score differences between the time-varying and static
    signed-dependence models across market states. Each point represents one
    held-out observation and plots the dynamic-minus-static Energy or Variogram
    Score against the VIX level or daily change in the
    10-year U.S. Treasury yield. Negative values indicate better performance
    of the time-varying specification. Solid lines are LOWESS smoothers and
    the horizontal dashed line denotes equal performance between the two
    models.}
    \label{fig:score_improvement_states}
\end{figure}

Figure~\ref{fig:score_improvement_states} examines whether the relative
held-out performance of the time-varying specification differs across market
states. The plotted values are the date-specific differences between the
dynamic and static scores, so observations below zero indicate that the
time-varying model provides a better fit on that date.

The LOWESS curves show that the advantage of the dynamic model is generally
maintained across the observed VIX range, particularly for the Variogram
Score, for which the smoothed score difference remains below zero over most
of the sample. The relationship between relative model performance and daily
changes in the 10-year Treasury yield is considerably weaker, with only
limited variation in the smoothed score differences across yield-change
states. This pattern is consistent with the estimated dynamic dependence
equation, in which the VIX has a substantially larger coefficient than the
10-year yield-change variable.

The futures application supports allowing the signed coupling parameter to vary with market conditions. In the static comparison, ordinary and scaled Dirichlet margins performed similarly. Allowing the coupling parameter to vary with the VIX and daily changes in the 10-year Treasury yield improved both the held-out energy score and the held-out variogram score. Several fitted cross-correlations differed in sign from their empirical counterparts, indicating that some component pairs exhibit dependence patterns beyond those captured by the single coupling parameter.

\appendix
\section{Additional Lemmas and Definitions}

\begin{lemma}[Dirichlet-average representation of Carlson's multiple
$R$-function \citet{JiangKadaneDickey1992,Dickey1983}]
\label{lem:carlson_multiple_R}

Let
\[
\mathbf Z
=
(Z_1,\ldots,Z_M)^\top
\sim
\operatorname{Dirichlet}_M(\boldsymbol{\gamma}),
\]
where $\boldsymbol{\gamma}
=
(\gamma_1,\ldots,\gamma_M)^\top$ for 
$\gamma_m>0$.
Let $\mathbf H
=
[\mathbf h_1,\ldots,\mathbf h_J]$
be an $M\times J$ matrix, and let $\boldsymbol{\nu}
=
(\nu_1,\ldots,\nu_J)^\top$
be such that the following expectation exists. Then the corresponding
Carlson multiple $R$-Dirichlet average may be written as
\[
\boxed{
\mathcal R
\left(
\boldsymbol{\gamma};
\mathbf H;
\boldsymbol{\nu}
\right)
=
\mathbb E_{\mathbf Z\sim
\operatorname{Dirichlet}(\boldsymbol{\gamma})}
\left[
\prod_{j=1}^{J}
\left(
\mathbf h_j^\top\mathbf Z
\right)^{\nu_j}
\right].
}
\]

Equivalently,
\[
\mathcal R
\left(
\boldsymbol{\gamma};
\mathbf H;
\boldsymbol{\nu}
\right)
=
\frac{1}{B(\boldsymbol{\gamma})}
\int_{\mathcal S^{M-1}}
\left\{
\prod_{j=1}^{J}
\left(
\mathbf h_j^\top\mathbf z
\right)^{\nu_j}
\right\}
\prod_{m=1}^{M}
z_m^{\gamma_m-1}
\,d\mathbf z,
\]
where $B(\boldsymbol{\gamma})
=
\frac{
\prod_{m=1}^{M}\Gamma(\gamma_m)
}{
\Gamma(\gamma_+)
}$ with $\gamma_+
=
\sum_{m=1}^{M}\gamma_m$.
\end{lemma}

\begin{remark}
Equation (1.2) of \citet{JiangKadaneDickey1992} expresses a special
case of Carlson's two-way multiple hypergeometric function as
\[
R(\mathbf b,\mathbf G,-\mathbf c)
=
\mathbb E_{\mathbf U\sim\operatorname{Dirichlet}(\mathbf b)}
\left[
\prod_{j=1}^{J}
\left(
\sum_{i=1}^{I}g_{ij}U_i
\right)^{c_j}
\right].
\]
The notation
$\mathcal R(\boldsymbol{\gamma};\mathbf H;\boldsymbol{\nu})$
is adopted here for convenience and to avoid
ambiguity arising from the sign convention in Carlson's
$R$-function notation.
\end{remark}

\bibliographystyle{unsrtnat}

\bibliography{main}

\end{document}